\pdfoutput=1
\documentclass[useAMS,usenatbib]{mnras}

\usepackage{graphicx}
\DeclareGraphicsExtensions{.pdf,.png,.jpg}
\usepackage{array}
\usepackage{hyperref}
\usepackage{pdflscape}

\long\def\symbolfootnote[#1]#2{\begingroup%
\def\thefootnote{\fnsymbol{footnote}}\footnote[#1]{#2}\endgroup} 
\usepackage{epsfig}
\usepackage{color}\DeclareGraphicsExtensions{.pdf,.png,.jpg}

\def\beq{\begin{equation}} 
\def\eneq{\end{equation}} \def\bea{\begin{eqnarray}} \def\enea{\end{eqnarray}}

\usepackage{natbib}
\usepackage[T1]{fontenc}
\usepackage{amsmath,amssymb}

\title[The URC of dwarf disc galaxies]{The universal rotation curve of dwarf disc galaxies}
\author[Karukes et al.]{E.V. Karukes$^{1,2,3}$\thanks{E-mail:
    ekarukes@sissa.it} P. Salucci$^{1,2}$\thanks{E-mail:
    paolo.salucci@sissa.it}\\$^{1}$SISSA International School for Advanced Studies, Via
  Bonomea 265, 34136, Trieste, Italy\\
$^{2}$INFN, Sezione di Trieste, Via Valerio 2, 34127, Trieste, Italy\\
$^{3}$ICTP South American Institute for Fundamental Research, and Instituto de F\'isica Te\'orica - \\
Universidade Estadual Paulista (UNESP), Rua Dr.~Bento Teobaldo Ferraz 271, 01140-070 S\~{a}o Paulo, SP Brazil
}

\date{Accepted 2016 November 22. Received 2016 November 4; in original form 2016 July 20}

\begin{document}

%\date{\today}

\pagerange{\pageref{firstpage}--\pageref{lastpage}} \pubyear{...}

\maketitle

\label{firstpage}

\begin{abstract}
We use the concept of the spiral rotation curves universality  to investigate the luminous
and dark matter properties of the dwarf disc galaxies in the local volume~(size~$\sim11$~Mpc).
Our sample includes 36 objects with rotation curves carefully selected
from the literature. We find that, despite the large variations of our sample in luminosities ($\sim$ 2 of dex), the rotation curves in specifically normalized units, look all alike and lead to the lower-mass version of the universal rotation curve of spiral galaxies found in Persic et al. We mass model the double normalized universal rotation curve $V(R/R_{opt})/V_{opt}$ of dwarf disc galaxies:  the results show that these systems are totally dominated by dark matter whose density shows a core size between 2 and 3 stellar disc scale lengths. Similar to galaxies of different Hubble types and luminosities, the core radius $r_0$ and the central density $\rho_0$ of the dark matter halo of these objects are related by $ \rho_0 r_0 \sim 100\hspace{0.1cm} \mathrm{M_\odot pc^{-2}}$. The structural properties of the dark and luminous matter emerge very well correlated. In addition, to describe these relations, we need to introduce a new parameter, measuring  the compactness of light distribution of a (dwarf) disc galaxy. These structural properties also indicate that there is no evidence of abrupt decline at the faint end of the baryonic to halo mass relation.  Finally, we find that the distributions of the stellar disc and its dark matter halo are closely related.
\end{abstract}   

\begin{keywords}

galaxies: dwarfs -- galaxies: formation -- galaxies: haloes -- galaxies: kinematics and dynamics -- dark matter.
\end{keywords}

\section{Introduction}
 
% \vspace{2.5cm}
% \vspace{2.7cm} 
 It is widely believed that only 15 per cent of the total matter in the Universe is in the form of ordinary baryonic matter. Instead the other 85 per cent
 is provided by dark matter (DM), which is detectable, up to now, only through its gravitational influence on luminous matter. The paradigm is that DM 
 is made by massive gravitationally interacting elementary particles with extremely weak, if not null interaction via other forces 
 \citep[e.g.,][]{white1982, jungman}. In this framework the well known  ($\Lambda$)CDM scenario, successfully describing the large structure of the Universe, has emerged \citep{kolb90}: accurate N-body simulations have
 found that the DM density profile of  the virialized structures such as galactic halos is universal and well described by the Navarro-Frenk-White profile \citep[hereafter NFW;][]{nfw}. 

However, at the galactic scales, this scenario has significant challenges. 

First, the apparent mismatch between  the number of the detected satellites around the Milky Way and  the predictions  of the corresponding simulations, known as the
"missing satellite problem" \citep{klypin99,moore99}, which also occurs  in the field galaxies \citep{zavala09,papastergis11,klypin15}. 
This discrepancy widens up when the masses  of the detected satellites are compared to those of the predicted subhalos  (i.e.  "too big to fail problem")  \citep[see][]{boylan-kolchin12,ferrero12,garrisonkimmel,papastergis15}.
 
Furthermore, there is the "core-cusp" controversy: the inner  DM density profiles of  galaxies generally  appear to be cored, and not cuspy as predicted 
 in the simplest ($\Lambda$)CDM scenario \citep[e.g.,][to name few]{salucci01,deblok02,gentile05,weinberg13, bosma04,simon05,gentile04,gentile07,donato09,delpopolo09,oh11}.
 
  These apparent discrepancies  between the observations and the predictions of the DM-only simulations suggest to either  abandon  the 
  ($\Lambda$)CDM  scenario  in favour of the others \citetext{e.g., selfinteracting DM \citealp{vogelsberger14,elbert15} or warm DM \citealp{devega13,devega133,lovell14,devega14}} or upgrade the role of baryonic 
  physics in the galaxy  formation process. The latter can be done including strong gas outflows, triggered by stellar and/or AGN feedback that are thought to strongly modify 
  the original  ($\Lambda$)CDM halo profiles out to a distance as large as the size of the stellar  disc \citep[e.g.,][]{navarro96,read05,
   mashchenko06,pontzen12,pontzen14,dicintio14}.

Although these issues are present in galaxies of any luminosity, however in low luminosity systems they emerge more clearly and appear much 
more difficult to be resolved within the  ($\Lambda$)CDM scenario. Thus, galaxies with I-band absolute magnitude $M_I \gtrsim -17$  play a pivotal role in that, observationally,  these  objects are dark matter dominated at all radii. Moreover in the ($\Lambda$)CDM scenario they may be related to the building blocks of more massive galaxies. In  light of this the importance of  dwarf spheroidal galaxies in various DM issues is well known \citep[see, e.g.,][]{gilmore07}. However, down to $M_{I} \sim -11$ there is no shortage of rotationally supported late-type systems, although a systematic investigation  is lacking. These rotationally supported systems have a rather simple kinematics suitable for investigating the properties of their dark matter content. 
 
In normal spirals, one  efficient way to represent and model their rotation curves (RCs)  comes  from the concept  of a universal rotation curve (URC). Let us stress that the concept of universality in RCs does not mean that  all of them have  a unique profile,  but that  all the RCs of $10^9$ local spirals (within $z \simeq 0.1$) can be  described by a same function of radius,  modulated  by few free parameters. They  depend  on the galaxy's global properties, namely  magnitude (or luminosity/mass) and a characteristic radius of the luminous matter\footnote{i.e. optical radius $R_{opt}$ defined as the radius encompassing 83 per cent of the total luminosity.} so that: $V(R)= V(R,L,R_{opt})$.
 
This concept, implicit in \citet{rubin85}, pioneered by \citet{persic91}, set by \citet{pss96} (PSS, Paper I) and extended to large galactocentric radii by \citet{salucci2007} has  provided us the mass distribution of (normal) disc galaxies in the magnitude range $-23.5 \lesssim M_I \lesssim -17$.\footnote{Extensions of the URC to other Hubble types are investigated in \citet{salucci97,noordermeer07}.} This curve $V(R,L,R_{opt})$, therefore, encodes all the main structural properties of the dark and luminous matter of every spiral \citep[PSS,][]{yegorova07}. In this paper, we work out to extend the RCs universality down  to low-mass systems and then, to use  it  to investigate the DM distribution in dwarf disc galaxies. 

Noticeably, for this population of galaxies the approach of stacking the available kinematics is very useful. In fact, presently, for  disc systems  with the optical velocity $V_{opt}\lesssim 61\hspace{0.1cm} km/s$,  some kinematical data have become  available (galaxies of higher velocities are investigated in the PSS sample). However, there are not enough individual {\it high quality high resolution extended} RCs to provide us with a solid knowledge of their internal distribution of mass. Instead,  we will prove that the 36 selected in literature {\it good quality good resolution reasonably extended} RCs (see below for these definitions), once coadded,  provide us with  a  reliable kinematics yielding to their  mass distribution.

In this work, we construct a sample of dwarf discs from the  local volume catalog (LVC) \citep[][hearafter K13]{karachentsev13}, which 
is $\sim 70$ per cent complete down to $M_{B}\approx-14$ and out to 11 Mpc, with the distances of galaxies obtained by means of  primary distance indicators.

Using LVC, we go  more than 3 magnitudes fainter  with respect to the sample of spirals of  PSS. Moreover the characteristics of the LVC guarantee us against several  luminosity biases that may affect such faint objects.    
 The total number of objects in this catalog is $\sim$900 of which $\sim180$ are dwarf spheroidal galaxies, $\sim500$ are dwarf disc galaxies and the rest are ellipticals and spirals. 
 
All our galaxies are low mass bulgeless  systems in which rotation, corrected for the pressure support, totally balances the gravitational force. Morphologically, they can be divided into two main types: gas-rich dwarfs that are forming stars at a relatively-low rate, named irregulars (Irrs) and starbursting
dwarfs that are forming stars at an unusually high rate, named blue compact dwarfs (BCD). The dwarf Irr galaxies are named "irregulars" due to the fact that they usually do not have a defined disc shape and
the star formation is not organized in spiral arms. However, some gas-rich dwarfs can have diffuse, broken spiral arms and be classified as late-type spirals (Sd) or as Magellanic spirals (Sm). The starbursting
dwarfs are classified as BCD due to their blue colours, high surface brightness and low luminosities. Notice that  it is not always easy to distinguish among  these types
since the galaxies we are considering often share the same parameters space for many structural properties \citep[e.g.,][]{kormendy85,binggeli94,tolstoy09}. 

In this paper, we neglect the morphology of the baryonic components as long as their stellar disc component follows a radially exponential surface density profile;  the identifiers of a galaxy are  $V_{opt}$, its disc length scale $R_D$ and its $K$-band  magnitude $M_K$  that can be substituted by  its disc mass.  We  refer to these systems of any  morphologies and $M_K \gtrsim -18 $ as dwarf discs (dd).

 In order to compare galaxy luminosities in different  bands, we write down  the  dd relations  between the  magnitudes in different bands $<B-K>\simeq2.35$ \citep{jarrett03} and $<B-I>\simeq1.35$ \citep{fukugita95}.

The plan of this paper as follows: in Section 2 we describe the sample that we are going to use; in Section 3 we introduce the analysis used to build
 the synthetic RC; in Section 4 we do the mass modelling of the synthetic RC; in Section 5 we denormalize 
 the results of the mass modelling in order to describe individually our sample of galaxies and then we define their scaling relations; in Section 6 we discuss our main results.

\section{The sample}  
 
We construct our dwarf disc galaxy sample  out of the  LVC \citep{karachentsev13}  by adopting the following 4 selection criteria:

1. We include disc galaxies with the optical velocity less than  $\sim 61$ $km/s$  (disc systems with larger velocity amplitude are studied in PSS);

2. The rotation curves  extend to at least 3.2 disc scale lengths.\footnote{$R_{opt}\simeq 3.2 R_D$} However, 
we allowed ourselves to extrapolate modestly the RCs of UGC1501, UGC8837, UGC5272 and IC10
due to their smoothness;

3. The RCs are symmetric, smooth and with an average internal uncertainty less than 20 per cent;

4. The galaxy disc length scale $R_D$  and the  inclination function $1/\mathrm{sin}\hspace{2px} i$  are   known  within   20 per cent  uncertainty.

 \begin{table}

\caption{Sample of dwarf disc galaxies. Columns: \textbf{(1)} galaxy name;  \textbf{(2)}  galaxy distance; \textbf{(3)} rotation curve source; \textbf{(4)}
 exponential scalelength of a galaxy stellar disc; \textbf{(5)} disc scalelenght source; \textbf{(6)} rotation velocity at the optical radius; \textbf{(7)} absolute magnitude in K-band.}
\label{tbl:1}
%\centering
 \begin{tabular}{lcccccc} 
	\hline
	 Name  &  D& RCs refs & $R_{ D}$ &  $R_{ D}$ refs  & $V_{ opt}$ & $M_K$\\ ---  & Mpc&---   &  kpc  &  --- &  km/s & mag \\
	 (1) & (2) & (3) & (4) & (5) & (6) & (7)\\\hline

	  &  & $H_{ \alpha}$;    HI  &  &     \\
    
      UGC1281 &4.94 &  1; 2   &   0.99 & a & 53.8&-17.97\\   
      UGC1501 & 4.97 &1; -&  1.32 &  a & 50.2&-18.19\\
      UGC5427 &  7.11 &1; -   &  0.38 &  e& 54.0&-17.06\\
%     DDO99 &   &$H_{\alpha} - 1$  &   0.27 &  c& 12.5\\
%     UGC4305 &  & $H_{I} - 2$, $H_{I} - 3$  &  1.11&  d& 34.5\\
      UGC7559 &4.88 & -;2, 3   &  0.88 &  b& 37.4&-16.91\\
      UGC8837 & 7.21 &-; 2  &   1.55 &  b & 47.6&-18.25\\
      UGC7047 &  4.31 &-; 2,4  &  0.57 &  c & 37.0&-17.41\\
      UGC5272 &  7.11 &-; 2  &   1.28 &  b & 55.0&-16.81\\
      DDO52 & 10.28 &-; 3   &   1.30 &  b & 60.0&-17.69\\
      DDO101 &16.1 &-; 3   &   0.94 &  b & 58.8&-19.01\\
      DDO154 &  4.04 &-; 3   &   0.75 &  b & 38.0&-15.70\\
      DDO168 &  4.33  &-; 3   &   0.83 &  b & 60.0&-17.07\\
      Haro29 &  5.68 &-; 3,4  &  0.28 &  b & 32.6&-16.26\\
      Haro36 & 8.9 &-; 3  &  0.97 & h & 56.5&-17.63\\
      IC10 &   0.66 &-; 3   &   0.38 &  b& 41.0&-17.59\\
      NGC2366 &3.19   &-; 3,4  &   1.28 &  b& 55.0&-18.38\\
      WLM &   0.97 &-; 3    &   0.55 &  b & 33.0&-15.93\\
      UGC7603 & 8.4 &-; 2   &   1.11 &  2 & 60.3&-19.07\\
      UGC7861 &  7.9 &-; 5  & 0.62 &  i& 61.0&-19.74\\
      NGC1560&  3.45 &-; 6   &   1.10 &  6& 56.1&-18.43\\
      DDO125 &   2.74 &1; 2   &   0.49 & c& 17.0&-16.96\\
%      DDO190 &  2.74 &-; 7   &   0.30 &  c & 22.2\\
%      KK149 &  8.9 &1; - &   0.34 &  g& 28.0\\
      UGC5423 & 8.71  &1; 12   &  0.52 &  d& 39.5&-17.71\\
%      UGC6456 &  & $H_{\alpha} - 1$, $H_{I} - 8$  &   0.51 &  b& 14.7\\
      UGC7866 &  4.57 &-; 2  &   0.54 &  2& 28.7&-17.18\\
      DDO43 &  5.73 &-; 3  &   0.57 & b& 35.3&-15.72\\
      IC1613 &  0.73 &-; 3   &   0.60 &  b& 19.0&-16.89\\
      UGC4483 & 3.21  &-; 4   &  0.16 & f& 20.8&-14.20\\
      KK246 & 7.83 &-; 9   &  0.58 & 9& 34.6&-16.17\\
%      UGC3476 &  & $H_{\alpha} - 1$   &   0.25 &  f& 9.0\\
      NGC6822 & 0.5 &-; 10   &   0.56 & b& 35.0&-17.50\\
      UGC7916 &   9.1 &-; 2    &  1.63 & h& 37.0&-16.22\\
      UGC5918 &  7.45 &-; 2   &   1.23 &  2& 45.0&-17.50\\
%      DDO133 &   &$H_{I} - 3$    &   1.71 &  b& 40.84\\
%      DDO216 &   &$H_{I} - 3$   &   0.37 & b& 13.25\\
      AndIV &  7.17 &-;11   &   0.48 &  11 & 32.2&-14.78\\
%       2.0 &   86.2   &   79.0 &  7.0& ---\\
	UGC7232 & 2.82  &-; 2   &   0.21 &  f & 37.0&-16.46\\
	DDO133 & 4.85  &-; 3    &  0.9 &  g& 42.4&-17.31\\
	UGC8508 & 2.69  &1; 3    &  0.28 &  j& 25.5&-15.58\\
        UGC2455& 7.8  &-; 2    &  1.06 &  h& 47.0&-19.91\\
        NGC3741& 3.03 & -; 7 &   0.18  & c &  23.6&-15.15\\
        UGC11583 & 5.89& -; 8  & 0.17& 8 &52.2&-16.55\\
        
                  \hline      
            \end{tabular}

            \textit{Notes.}
 %, Simpson et al. (2012)-8,  
 %, Leroy et al. (2008)-d 4305   
 %\citet{bershady10}-e,      
RC $\&$  $R_{ D}$  references: \citet{moiseev14}-1, \citet{swaters09}-2, \citet{oh15}-3, \citet{lelli14}-4, \citet{epinat08}-5, \citet{gentile10}-6, \citet{gentile07}-7, \citet{begum04}-8; \citet{gentile12}-9, \citet{weldrake03}-10, \citet{karachentsev16}-11,\citet{oh11}-12, \citet{vanzee01}-a, \citet{hunter04}-b, \citet{sharina08}-c,  \citet{parodi02}-d,
  \citet{simard11}-e, \citet{martin1998}-f, \citet{hunter11}-g, \citet{herrmann13}-h, \citet{yoshino15}-i, \citet{hunter12}-j. Distance $D$ and absolute magnitude in K-band $M_K$ are taken from \citet{karachentsev13}.

\end{table}

It is worth noticing that  for an RC to fulfil  criteria (2), (3) and (4) it  is sufficient to qualify it for the  {\it coaddition} procedure  but not necessarily  this is the case for  the {\it individual}  modelling.

The kinematical data used in our analysis are HI   and $H_{\alpha}$ rotation curves available in the literature (see Table \ref{tbl:1}), which are corrected for inclination and instrumental effects. Furthermore, circular velocities of low mass galaxies, with $V_{max} \lesssim 50 \hspace{2px}  km/s$, require to be checked for the pressure support correction, this can be done using  the so-called "asymmetric drift correction" \citep{dalcanton10}. Therefore, most of the RCs in our sample either have the asymmetric drift correction applied \citep[the ones taken from][]{oh11,oh15,lelli14,gentile10,gentile12} or pressure support has been determined and is too small to affect the RC \citep[the ones taking from][]{swaters09,weldrake03,karachentsev16}. Despite that, we leave three galaxies (UGC1501, UGC5427 and UGC7861) for which circular velocities were not corrected. In view of their $V_{max}$ are larger than 50 $km/s$, therefore the effect should be minor. In the innermost regions of galaxies, when available, we use also $H_{\alpha}$ data not corrected for the asymmetric drift since such term is negligible  as it was pointed out by e.g. \citet{swaters09,lelli12}.

We stress  that the above selection process has left out  few objects  whose RCs are sometimes considered in literature, e.g. the RC of DDO 70 described by \citet{oh15} has failed our criteria (3) because of its abnormal shape. Our approach stands firmly that, in order to provide us with proper and correct information about  a galaxy dark and luminous mass distribution, the relative kinematical and the photometric data must reach a well defined level of quality, otherwise they will be confusing rather than enlightening the issue.

 Therefore, we ended with the final sample of 36 galaxies spanning the magnitude and  disc radii intervals,  $-19 \lesssim~M_{I} \lesssim -13$, $0.18 \hspace{2px} kpc \lesssim R_{D} \lesssim 1.63 \hspace{2px} kpc$ and the optical
 velocities  $17 \hspace{2px} km/s\lesssim V_{opt} \lesssim 61 \hspace{2px} km/s$  (see Table \ref{tbl:1},
 the references for HI and H$\alpha$ RCs of our sample are also given in the table and the individual RCs are plotted in Fig.\ref{figa1} of Appendix~A). 
 
 The average optical
radius $ \langle R_{opt} \rangle$ and the $\log$ average optical velocity $ \langle V_{opt} \rangle$ of our sample are $2.5 \hspace{2px}  kpc$, $40.0 \hspace{2px}  km/s$, respectively, (these values will be specifically  needed  in  Section~5).
For a comparison, the galaxy sample of PSS spans the magnitude interval $ -23.5 \lesssim M_I \lesssim -17$,  
 the optical disc radii $ 6.4 \hspace{2px} kpc \lesssim R_{D} \lesssim 96 \hspace{2px} kpc$, and the optical velocities between $70 \hspace{2px}  km/s\lesssim V_{opt} \lesssim300 \hspace{2px}  km/s$. 
 Therefore, our sample will extend the URC of PSS by 2 orders of magnitude down in galaxy luminosity. However, let us stress that, differently from PSS, we are going to construct only one luminosity bin. This is, firstly, due to the fact that the amount of galaxies in our sample is small compare to that of PSS and, secondly, due to the fact that, after the normalising procedure, they all converge to the same RC profile independently of the galaxy luminosity (see next section).

   \begin{figure*}
 \begin{centering}
\includegraphics[angle=0,height=6.5truecm,width=18truecm]{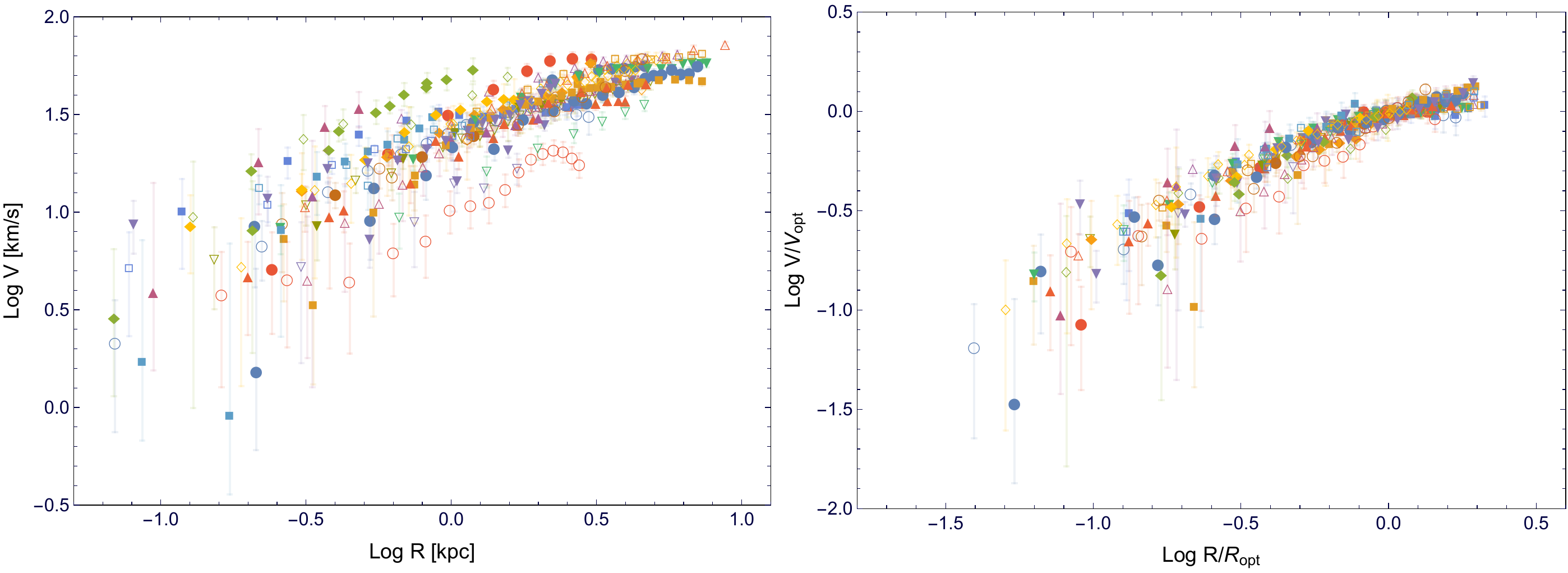}
\caption{Individual RCs. {\it Left panel:} in physical units. {\it Right panel:} after double normalization on $R_{opt}$ and $V_{opt}$.}
\label{fig1}	
\end{centering}
\end{figure*}

 \begin{figure*}
 \begin{centering}
\includegraphics[angle=0,height=7.truecm,width=9truecm]{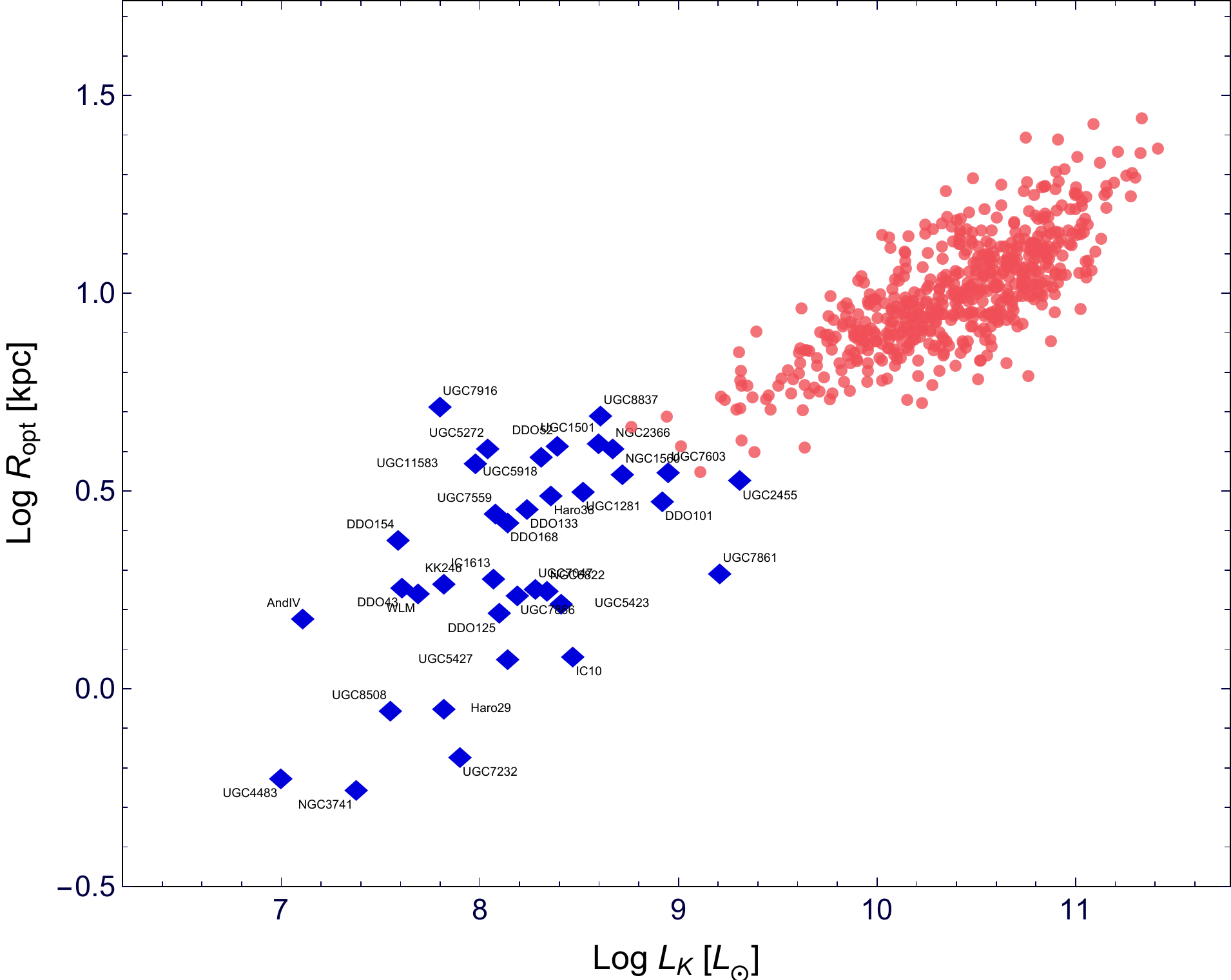}
\caption{The optical radius versus the total luminosity. Red circles indicate the normal spiral galaxies from the sample of PSS and blue diamonds are the dwarf galaxies of the present work.}
\label{fig2}	
\end{centering}
\end{figure*}

\section{The URC of Dwarf discs}

\begin{figure*}
 \begin{centering}
\includegraphics[angle=0,height=12truecm,width=16truecm]{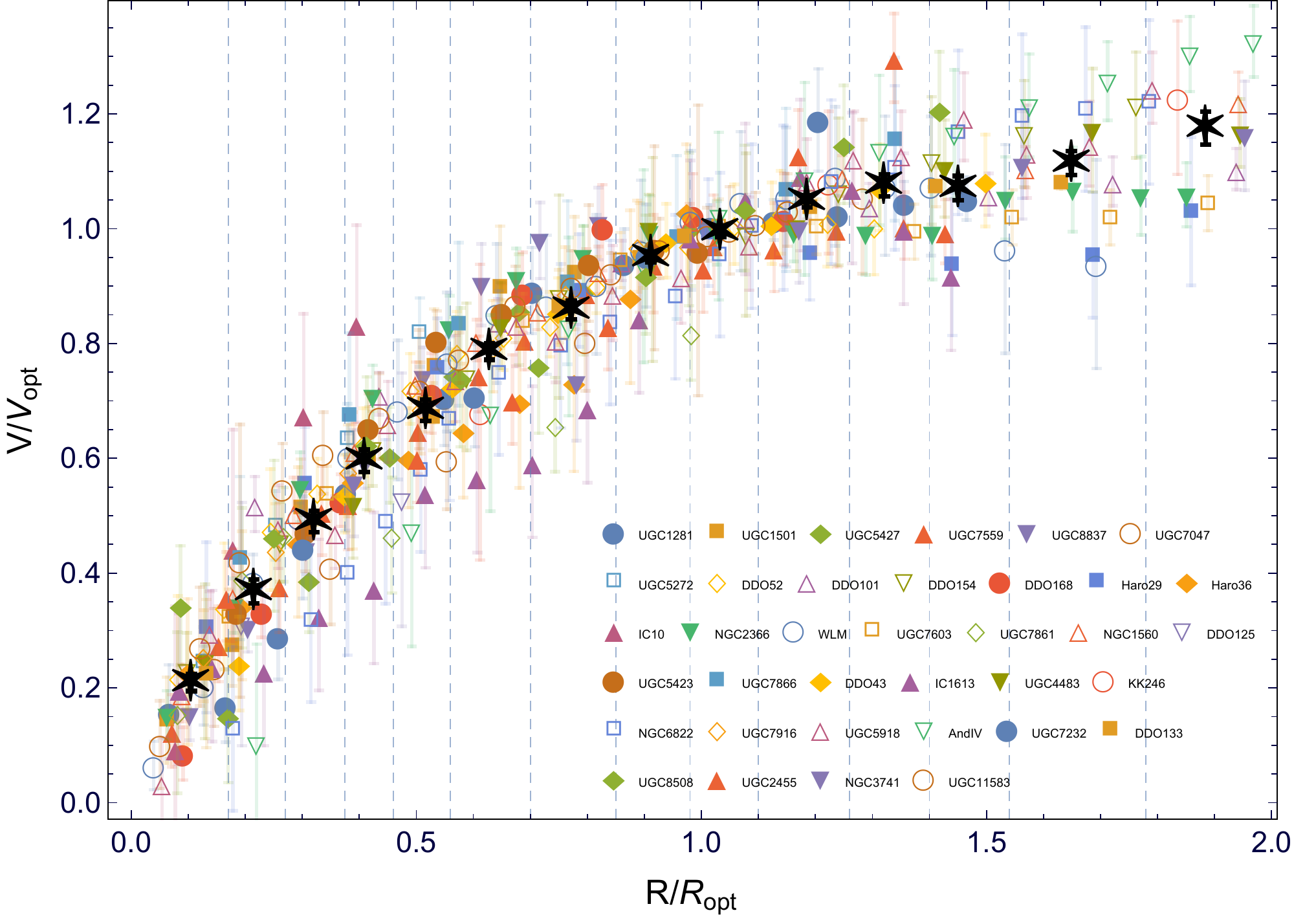}
\caption{Individual RCs normalized to $R_{opt}, V_{opt}$. Black stars indicate the synthetic RC. Bins are shown as vertical dashed grey lines. }
\label{fig3}	
\end{centering}
\end{figure*}

\begin{table}
\caption{Data in the radial bins. Columns: \textbf{(1)} bin number;  \textbf{(2)} number of data points; \textbf{(3)} the central value of a bin; \textbf{(4)} 
the average coadded weighted normalized rotation velocity; \textbf{(5)} rms on the average coadded rotation velocity \textbf{(6)} denormalized on  $\langle R_{opt} \rangle$  values of radii, $kpc$ \textbf{(7)} denormalized on  $\langle V_{opt} \rangle$  values of velocities, $km/s$ \textbf{(8)}  rms on the denormalized rotation velocity.  \label{tbl:2}} 
%\label{tbl:2}

%\centering

  \begin{tabular}{ l c c c c c c c c} 

	\hline
	i&  N & $r_{i}$ & $v_{i}$  &$d{v_{i}}$ &$R_i$ &  $V_{i}$ & $d{V_{i}}$\\

	(1) & (2) & (3) & (4) & (5) & (6) & (7) & (8)\\\hline\\
	 %\\ ---  & ---& kpc   &  km/s  & km/s  

     1 & 31 &  0.11  &0.21  & 0.015  & 0.27  &  8.38 &  0.60\\   
      2 & 30 &0.22    &0.37 & 0.021 & 0.54 & 14.57 & 0.83\\
      3 &  21 &0.32    &0.49& 0.019 & 0.81 & 19.61 & 075\\
      4 & 26  &0.41     &0.60&  0.019  & 1.03  & 23. 90 &  0.78\\
      5 & 25  &0.52    &0.68 &  0.018 &  1.30  & 27.36 & 0.74\\
      6 &  33  &0.63     &0.78 &  0.014 & 1.58 & 31.41 & 0.57\\
      7 &  34 &0.77    &0.86  &  0.016 &  1.94 & 34.31 & 0.65\\
      8 & 28 & 0.91    & 0.95 &  0.009 &  2.29 & 37.88 & 0.35\\
      9 & 25 &1.03     &0.99  &  0.009 &  2.60 & 39.64 & 0.38\\
      10 & 28 &1.18   &1.05 &  0.010 &  2.97  &41.79 & 0.39\\
      11 & 18&1.32     &1.07  &  0.018 & 3.31 & 42.97& 0.72\\
      12 & 17&1.45  &1.07  &  0.020 & 3.65 & 42.68 & 0.78\\
      13 & 20&1.65  &1.12  &  0.020  & 4.13 & 44.70 & 0.80\\
       14 & 14&1.88  &1.20 &  0.030  & 4.73 & 47.83  &  1.18\\

                  \hline
\end{tabular}
\end{table}

First, we plot the RCs of galaxies in our sample expressed in physical units in log-log scales (see the left panel of Fig.~\ref{fig1}). We realize, even at a first glance, that, contrary to the RCs of normal spirals  \citep[see, e.g.,][PSS]{yegorova07}, each dd galaxy has an RC with a very different profile, as it has also been noticed  by \citet{oman15}. In other words, all curves are rising with radius but at a very different place. 

  Surprisingly, the origin of such diversity is closely linked with the very large scatter  that the dds show in  the relationship between the optical radius $R_{opt}$ and  the luminosity  $L_K$ shown in  Fig. \ref{fig2}. In our sample the relation still holds but the scatter remarkably increased, while in normal spirals luminosities and disc sizes are very well correlated.

Thus, in dwarf disc galaxies, by following the analogous PSS procedure, we are going to derive a universal profile of their RCs.
As an initial step of the co-addition procedure (see PSS for  details) each of the 36 $V(R)$ has been normalized to its own optical radius $R_{opt}$ and to its optical velocity $V_{opt}$ obtained by RC data interpolation. We then derive the quantity $V(R/R_{opt})/V(R_{opt})$. This double normalization  
  eliminates most of the small scale individualities of the RCs.\footnote{The double normalization refers to both quantities plotted on the X axis and Y axis of Fig. \ref{fig3}.} In the right panel of Fig. \ref{fig1} you can see that all the double normalized RCs of our sample converge to a profile very similar to that of the least luminous normal  spirals (red joined circles of Fig. \ref{fig3a}).

Notice, that this effect is not new: in \citet{Verheijen99,salucci97} \citep[see also][]{mcgaugh14} the variety of RCs shapes in physical units between high surface brightness galaxies and LSB objects of similar maximum velocities were eliminated by  normalizing $V(R)$ on the corresponding disc scale lengths. Related to this issue, there are also several studies that have analysed quantitatively the shapes of the RCs of different morphological types of galaxies \citep[see, e.g.,][]{swaters09,lelli14,errozferrer16}.

 \begin{figure*}
\begin{centering}
\includegraphics[angle=0,height=7truecm,width=10truecm]{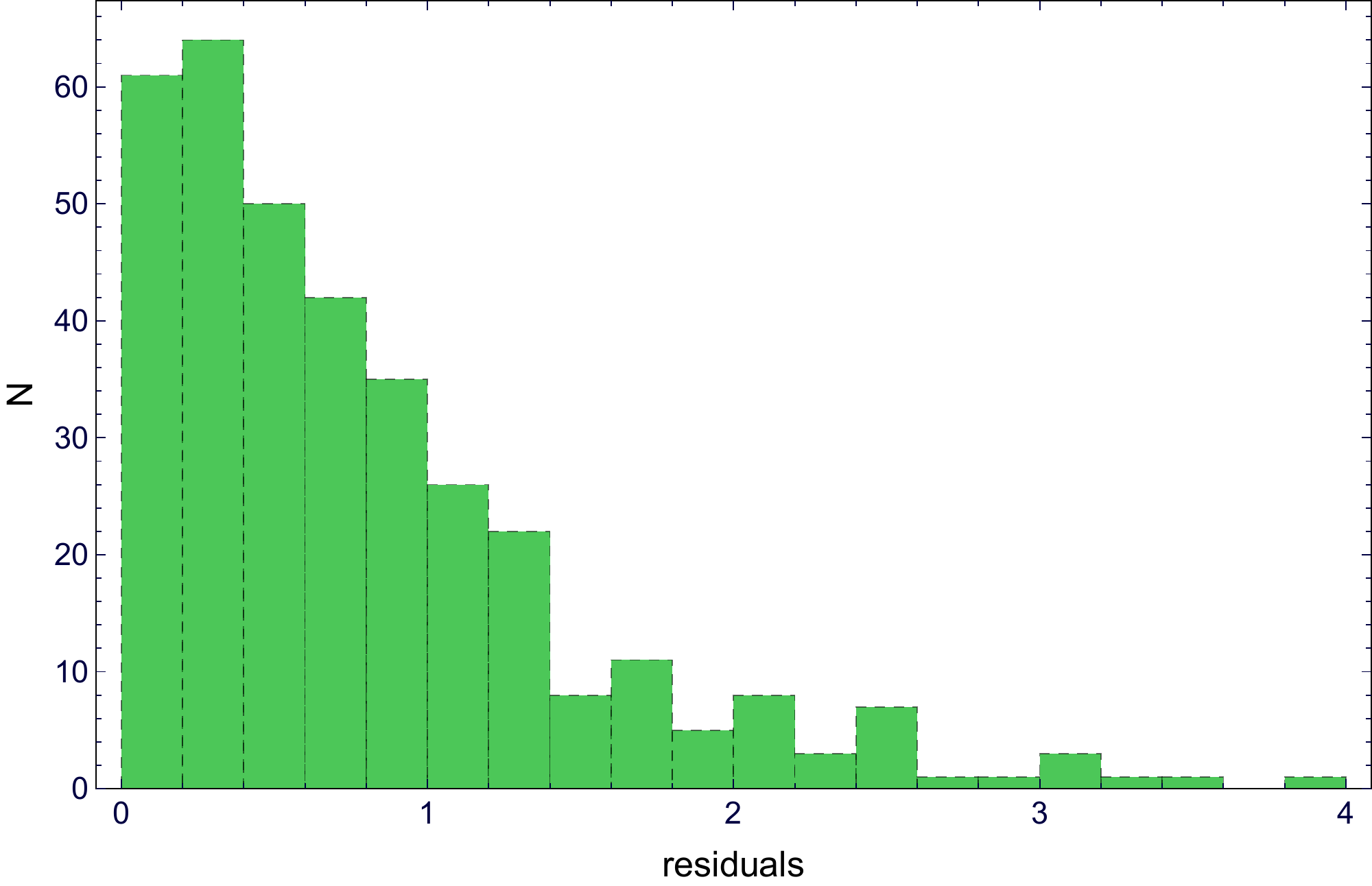}
\caption{The distribution of residuals in terms of rms, which are listed in column 5 of Table \ref{tbl:2}.}
\label{fig4}	
\end{centering}
\end{figure*}

 \begin{figure*}
 \begin{centering}
\includegraphics[angle=0,height=7truecm,width=10truecm]{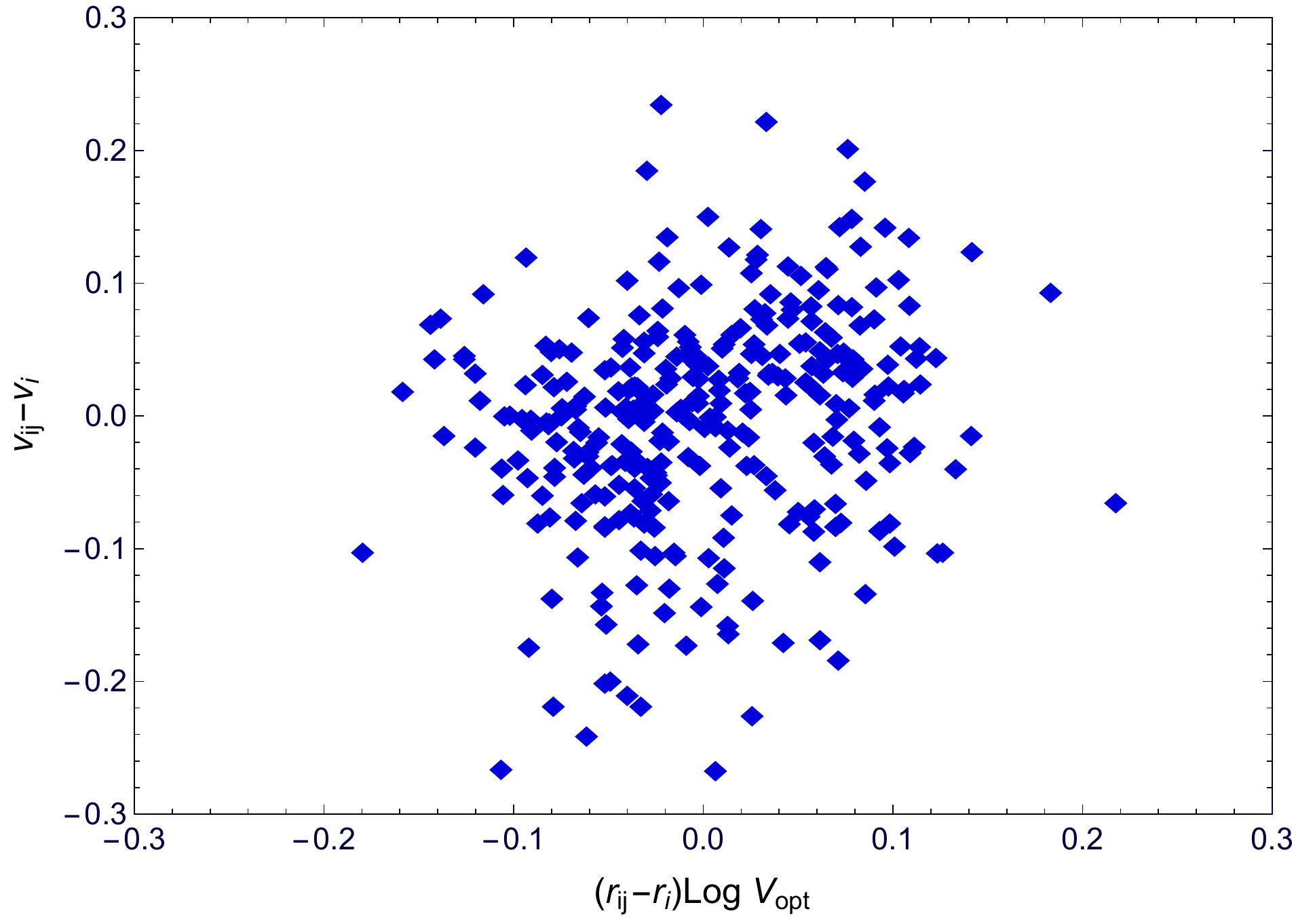}
\caption{The $v_{ij}$ residuals in the 14 bins versus the optical velocity  $\log V_{opt}$. Coefficient of correlation $R^2$ is $\sim 0.06$.}
\label{fig44}	
\end{centering}
\end{figure*}

%\red{Let us recall, that in investigating the RC {\it profiles} of our sample we construct just one luminosity bin, because of the data limitation. At its support, let us note that the URC at low luminosities converges to a profile independently of luminosity.}
%\vspace{10px}

Next step is to obtain the corresponding raw synthetic RC. For this purpose the double normalized velocity data are co-added as follows: we set 14 radial bins in the position $r_i$\footnote{Calculated as a mean value in a bin.} given by the vertical dashed
 grey lines in Fig. \ref{fig3} and reported in Table~\ref{tbl:2}. Each bin is equally divided in two, we adopt that every RC can contribute to each of the 28 semibins only with one data point. For an RC with more data points concurring to the same semibin, these are averaged
accordingly. The last bin is set at  $r_i =1.9$ due to the lack of outer data.

Since a galaxy cannot contribute more than twice to every bin $i$, each of them, centred at
$r_i$ (see Table \ref{tbl:2}) and with boundaries shown in Fig. \ref{fig3}, has a number $N_i$ data
(see Table \ref{tbl:2}) which runs from a maximum of 68 and a minimum that we have set to be 14. Then, from $N_i$ data in each radial bin $i$ we compute the average weighted rotation velocity, where the weights are taken from the uncertainties in the rotation velocities (given online).

In Table \ref{tbl:2} we report the 14 $r_i$ positions, the values of the coadded double normalized curve $v =  V(R/R_{opt})/V(R_{opt})$  and  of their uncertainties $dv$, calculated as the standard deviation with respect to the mean.\footnote{Lowercase letters refer to normalized values, while capital letters to the values in physical units.} The universality of this curve can be inferred from its very small rms values  (see Fig. \ref{fig3}). Furthermore, we investigate the universality  in deep by
calculating the residuals of each individual RC with respect to the emerging coadded curve (Table \ref{tbl:2} column 5):

\begin{equation}
\begin{aligned}
\chi^2=\frac{\sum_{ij}\frac{(v_{ij}-v_i)^2}{\delta_{ij}^2}}{N},
\end{aligned}
\end{equation}

 \noindent
where $v_{ij}$ are the individual RC data  referring to the bin $i$ of the double normalized RC of the $j$ dwarf discs ($j$ from 1 to 36), $ v_{i}$ is the double normalized coadded $i$ value  ($i$ from 1 to 14) (see Fig. \ref{fig3}), $\delta_{ij}$ are the
 observational errors of the normalized circular velocities and $N$ is the total number of the data points ($N=350$).\footnote{$i$ is the index number of a radial  bin and $i j$ are the index numbers of a data $j$ in a bin $i$.}

  We consider the 14 $v_i$ as an exact numerical function which we attempt to fit with double normalized velocity data $v_{ij}$: we found  that  the fit is excellent, the  reduced chi-square is $\sim 1.0$ and the reduced residuals $dv_{ij} = v_{ij}-v_i$ are very small, see Fig. \ref{fig4}. In fact  $\sim 72$ per cent of the residuals is smaller than  1$~\delta_{ij}$,  $\sim 26$ per cent  falls inside 3~$\delta_{ij}$ and only the remaining $\sim 2$ per cent is anomalously large.   
 
 Finally, in order to check the existence of  biases, we investigate, in all 14 bins,  whether  the $dv_{ij}$ residuals have any correlation with the optical velocity  $\log V^{ij}_{opt}$ of the corresponding galaxy (see Fig. \ref{fig44}). However, we did not find any correlation, see Fig. \ref{fig44}; dd galaxies of any luminosity (and $V_{opt}$) show the same double normalized RC profile. Indeed, our accurate analysis shows no evidence of  dd with a double normalized RC to be different from the coadded $v_i(r_i)$ derived in this section and  given in Table \ref{tbl:2}.

Hereafter, it is worth comparing the present results  with those of  PSS; in the left panel of Fig \ref{fig3a}    the former is plotted alongside with the similar curves of the four PSS's luminosity  bins (see also Fig. 6 in PSS). The least luminous bin of PSS (red joint  circles in Fig \ref{fig3a}) contains 40 normal spirals with the average I-band magnitude of $\langle M_{I} \rangle=-18.5$. Noticeably, the two double normalized coadded  RCs are in good agreement, keeping in mind that, in the present work, the luminosity bin  $-19 \lesssim M_{I} \lesssim -13$ is as big as the whole luminosity range of PSS that, instead, was divided in 11 bins. 

Thus, starting from $M_I\sim -18.5$ and down to the faintest systems, the mass structure of disc galaxies is just a dark halo with a density core radius as big as the stellar disc. At $ M_{I}  \gtrsim-18.5$ the stellar disc contribution  disappears  and,  remarkably, the RC profile becomes solid body like:  $V(R)\propto R$.

We now investigate quantitatively the last statement:  one can notice that the  coadded RC of  dwarf discs is slightly shallower  than that of the least luminous  spirals of PSS (see left panel of Fig. \ref{fig3a} and Appendix B). Therefore, we check for the presence of  any trend  between  luminosity and shape of  the corresponding  RC inside our sample of 36 dwarf discs. We divide them  in 3 subsamples (12 galaxies  each, ordered by their luminosity). Then, we  derive the  3 corresponding stacked RCs (see Table \ref{tbl:1app}). No trend between RC shape and  luminosity was found, differently from what it occurs for spirals of higher luminosity, see  Appendix B and Fig \ref{fig3a}.

Finally, let us  point out  that neither the double normalization, nor the stacking of our 36 objects  is the cause of  the solid body profile of the RC in Fig.~\ref{fig3}  and Table \ref{tbl:2}.
The reason is that each RC of our sample, also when considered  in physical units, shows, inside $2 R_D$, a solid body profile

Therefore, we conclude the existence of the coadded RC for the dd population. This is the first step to obtain  that  the kinematics of  dd galaxies can be described by a  smooth universal  function,  exactly as it happens in normal spirals \citep[PSS,][]{salucci2007}.

\section{Modelling the double normalized coadded RC of dwarf disc galaxies}

As in normal spirals (see PSS) we mass model the  coadded RC data that represent the whole  kinematics  of dds. More precisely, 

1. the  coadded (double normalized) RC (see Table \ref{tbl:2}), once proven to be universal,  is  the basic data with which we build the mass model  of  dwarf disc galaxies; 

2. for simplicity, we rescale the 14 normalized velocities  $v_i$  to the average values of the sample  $ \langle V_{opt}\rangle$ and $\langle R_{opt}\rangle$, 40.0 km/s and 2.5 kpc, respectively. In details, we write:

\begin{eqnarray}
\langle V_{i}\rangle=v_{i} \langle V_{opt}\rangle;\nonumber\\
\langle R_{i}\rangle=r_{i} \langle R_{opt}\rangle,
\label{scaleopt}
\end{eqnarray}

\noindent
the 14 values of $\langle V_{i} \rangle$  and $\langle R_{i} \rangle$ are also reported in Table \ref{tbl:2} (columns 6-7), where angle brackets indicate normalization to the average values of optical radius and to the log average values of optical velocity. This RC is the fiducial curve  for dwarf disc systems. In fact,  we take the coadded curve  Table \ref{tbl:2} (columns 3-5) and we apply it to a galaxy  with the values of $R_{opt}$  and $V_{opt}$ equal to the average values in our sample. Since all dd RCs have the same double normalized profile, the parameters of the resulting mass model  can be easily rescaled back to cope with galaxies of different $V_{opt}$ and  $R_{opt}$.

 The fiducial rotation  curve (Table \ref{tbl:2} columns 6-8) of dwarf discs consists of 14 velocity data points extended out to 1.9 $\langle R_{opt}\rangle$. The uncertainties on the velocities  are at the level of $\sim 3$ per cent (see Fig \ref{fig3}).

Then the circular  velocity model  $V_{tot}(R)$  consists  into the sum, in quadrature, of three terms $V_{D}, V_{HI}, V_{DM}$  that describe  the contribution from  the stellar disc ,the HI disc and dark halo and that must equate to the  observed circular velocity:

\begin{equation}
V^2_{tot}(R)=V_m(R)\equiv V^2_{D}(R)+V^2_{HI}(R)+V^2_{DM}(R).
\label{circvel}
\end{equation}

Notice that in the RHS of eq. \ref{circvel}, we have neglected the stellar bulge contribution that is, in fact, absent in dds.

 \begin{figure*}
 \begin{centering}
\includegraphics[angle=0,height=5truecm,width=8truecm]{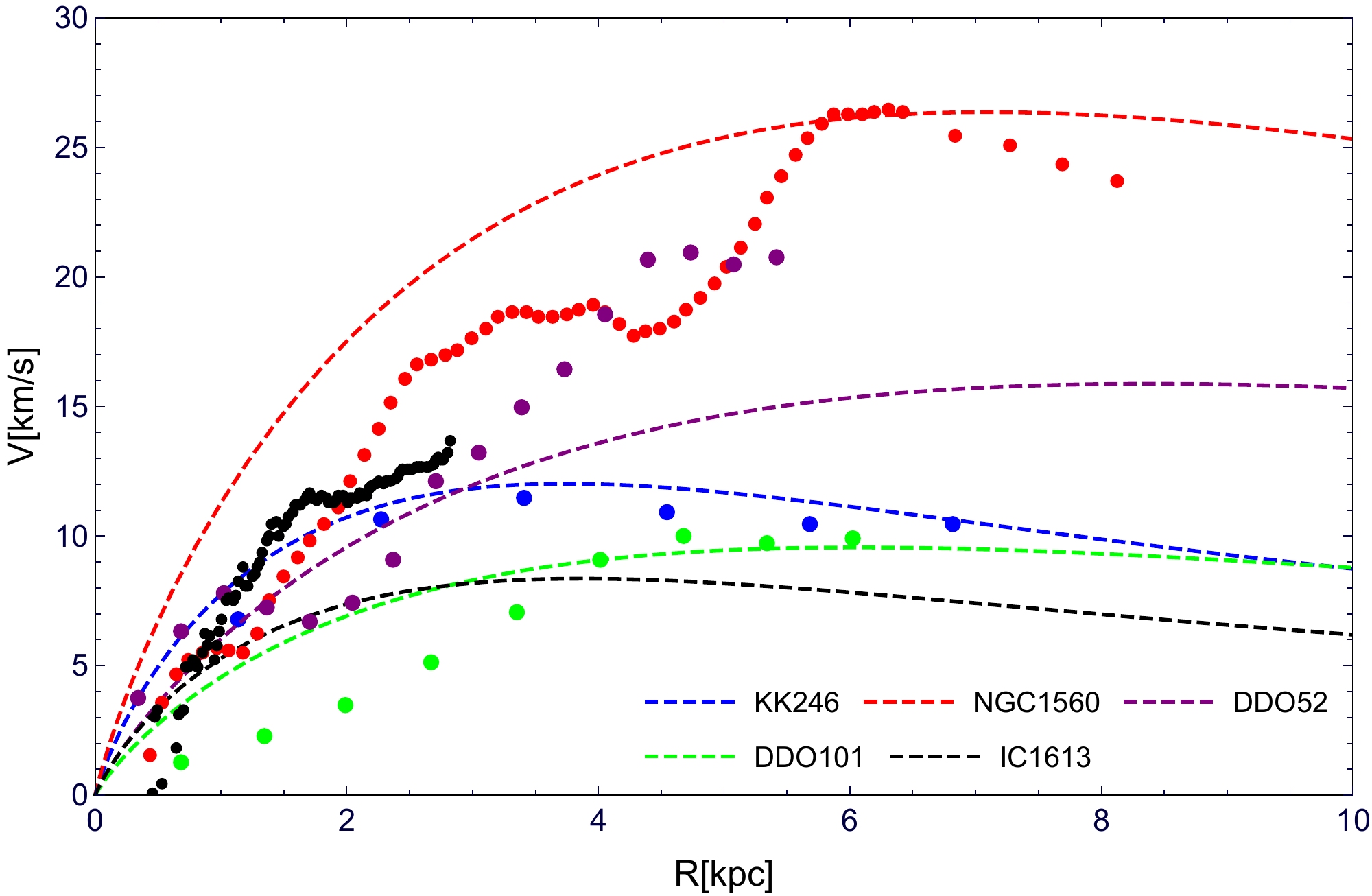}
\caption{The observed circular velocities of HI taking from \citet{gentile10, gentile12, oh15} (points) and the approximation for the distribution of HI component as described in \citet{tonini06} (dashed lines).}
\label{fig55}	
\end{centering}
\end{figure*}  

\subsection{Stellar component}

With a constant stellar mass to light ratio as function of radius \citep[see e.g][]{bell01} all 36 dds have the same surface density stellar profile  $\Sigma _{D}$ well represented by the Freeman disc \citep{freeman}:

\begin{equation}
\Sigma _{D}(R)=\frac{M_{ D}}{2 \pi R^2_{ D}}e^{-\frac{R}{R_{ D}}}
\label{freemandiscsigma}
\end{equation}  

\noindent
then, the contribution of the stellar disc  $V_{D}(R)$ is

\begin{equation}
V^2_{D}(R)=\frac{1}{2}\frac{G M_{D}}{R_{D}} (3.2 x)^2 (I_{0} K_{0}-I_{1} K_{1}), 
\label{freemandisc}
\end{equation}
 
 \noindent
where $x=R/R_{opt}$ and $I_{n}$ and $K_{n}$ are the modified Bessel functions computed at 1.6 x.

\subsection{Gas disc}

For each galaxy the gaseous mass $M_{HI}$ was taken from K13, log averaged and then multiplied by a factor 1.33 to account for the He abundance, then we obtain  $\langle M_{HI} \rangle = 1.7 \times 10^8 M_{\odot}$.
The HI surface density profile is not available for all dwarf disc galaxies of our sample, therefore we model it, by
following  \citet{tonini06}, as a Freeman distribution  with a scale length three times larger  that of the stellar disc $\Sigma _{HI}(R)=\frac{M_{HI}}{2 \pi (3 R_{D})^2}e^{-\frac{R}{3 R_{ D}}}$. Then the contribution of the gaseous disc  $V_{HI}(R)$ is

\begin{equation}
V^2_{HI}(R)=\frac{1}{2}\frac{G M_{HI}}{3 R_{D}} (1.1 x)^2 (I_{0} K_{0}-I_{1} K_{1}), 
\label{freemandiscgas}
\end{equation}

\noindent
where $x=R/R_{opt}$ and $I_{n}$ and $K_{n}$ are the modified Bessel functions computed at 0.53 x.

 This scheme is fairly well supported in dds for which  the HI surface density data  are available \citep[e.g., data from][]{oh15,gentile10,gentile12}. In order to clarify the latter, we plot  in Fig. \ref{fig55}, alongside the observed RC of HI component and our approximation of the HI distribution for 5 galaxies of our sample.
 
 In addition, let us stress that the gas contribution is always a minor component to the dds circular velocities, consequently possible errors in its estimate do not alter the mass modelling neither affect any result of this paper.

\subsection{Dark halo}

Many different halo radial mass profiles have been proposed over the years. Thus, in this work we are going to test the following profiles.

\subsubsection{Burkert profile}

The URC of normal spirals and the kinematics of individual objects  \citep{salucci2000} point  to dark halos density profiles with a constant core, and, in particular, to the Burkert halo profile \citep{burkert95}, for which:

\begin{eqnarray}
\rho_{B,URC}(r)=\frac{\rho_{0} r_{c}^3}{(r+r_{c})(r^2+r_{c}^2)},
\label{urc}
\end{eqnarray}

\noindent
where $\rho_{0}$ (the central density) and $r_{c}$ (the core radius) are the two free parameters and $\rho_{B,URC}$ means that we have adopted the Burkert profile for the URC DM halo component. Hereafter, we will freely exchange the two denominations according to the issue considered.

Adopting spherical symmetry, the mass distribution of the Burkert halos is given by:

\begin{equation}
 M_{URC}(r)= 2 \pi \rho_{0} r_{c}^3  \big[ln\big(1+\frac{r}{r_{c}}\big)-tg^{-1}\big(\frac{r}{r_{c}}\big)+0.5ln\big(1+\big(\frac{r}{r_{c}}\big)^2\big)\big]
 \label{burmass}
\end{equation}

\subsubsection{NFW profile}

We will investigate also NFW profile. \citet{nfw} found, in numerical simulations performed in the ($\Lambda$)CDM scenario of structure formation, that
 virialized systems follow a universal dark matter halo profile.  This is written as:

\begin{equation}
\rho_{NFW}(r)=\frac{\rho_{0}}{(\frac{r}{r_{s}})(1+\frac{r}{r_{s}})^2},
\label{nfw}
\end{equation}

\noindent
where $\rho_{0}$ and $r_{s}$ are, respectively, the characteristic density and the scale radius of the distribution. These two parameters can be
 expressed in terms of the virial mass $M_{vir}=4/3 \pi 100 \rho_{crit} R^3_{vir}$, the concentration parameter $c=\frac{R_{vir}}{r_{s}}$ and the critical density of the Universe
  $\rho_{crit}=9.3 \times 10^{-30} {g \ cm^{-3}}$. By using eq.(\ref{nfw}), we can write:

\begin{eqnarray}
\rho_{0} &=& \frac{100}{3} \frac{c^3}{\log{(1+c)}-\frac{c}{1+c}} \ \rho_{crit}\quad {g\,cm^{-3}};\nonumber\\
r_{s} &=& \frac{1}{c}\left(\frac{3 \times M_{vir}}{4 \pi 100 \rho_{crit}}\right)^{1/3} \quad {kpc}.
\label{rho}
\end{eqnarray} 

Then, the RC curve for the NFW DM profile is

\begin{equation}
V^2_{NFW}(r)=V^2_{vir}\frac{log(1+cx)-cx/(1+cx)}{x[log(1+c)-c/(1+c)]},
\label{nfw2}
\end{equation}

\noindent
where $x=r/R_{vir}$ and $V_{vir}$ represents the circular velocity at $R_{vir}$.

\vspace{0.3cm}

Let us point out that, within the ($\Lambda$)CDM scenario, the NWF profile maybe not the present days dark  halos around spirals. Baryons, during the formation  of the stellar discs, may have been able to  modify the original DM density distributions \citep[see, e.g.,][]{pontzen12,pontzen14,dicintio14}.
 We then consider eq. (\ref{nfw2}) as the fiducial profile of ($\Lambda$)CDM scenario, a working assumption useful to frame  changes of the latter.

\subsubsection{DC14 profile}

A solution for the existence of cored profiles in ($\Lambda$)CDM scenario  may have emerged by considering the recently developed DM density profile \citep[see][]{dicintio14}. This profile (hereinafter referred to as DC14) accounts for the effects of feedback on the DM halos due to gas outflows generated in high density starforming regions during the history of the stellar disc. The resulting radial profile is far from simple, since it starts from an ($\alpha, \beta,\gamma$) double power-law model \citep[see][]{dicintio14}

\begin{eqnarray}
\rho_{DC14}(r)=\frac{\rho_s}{(\frac{r}{r_s})^{\gamma}\left(1+(\frac{r}{r_s})^{\alpha}\right)^\frac{(\beta-\gamma)}{\alpha}}
\label{DC14}
\end{eqnarray}
\\
where $\rho_s$ is the scale density and $r_s$ the scale radius. The inner and the outer regions have logarithmic slopes $-\gamma$ and $-\beta$, respectively, and $\alpha$ indicates the sharpness of the transition. These three parameters are fully constrained in terms of the stellar-to-halo mass ratio as shown in \citet{dicintio14}:

\begin{eqnarray}
\alpha&=&2.94-{log}_{10}\lbrack(10^{X+2.33})^{-1.08}+(10^{X+2.33})^{2.29}\rbrack\nonumber\\
\beta&=&4.23+1.34X+0.26X^2\nonumber\\
\gamma&=&-0.06+{log}_{10}\lbrack(10^{X+2.56})^{-0.68}+(10^{X+2.56})\rbrack
\label{slopes}
\end{eqnarray}
 
 where $X={log}_{10}\left(\frac{M_{D}}{M_{halo}}\right)$.

Then, using the definition of the enclosed mass, we can write down the expression for the scale density of the DC14 profile:

\begin{eqnarray}
\rho_{s}=M_{vir}/4 \pi \int\limits_0^{R_{vir}}\frac{r^2}{(\frac{r}{r_s})^\gamma\lbrack1+(\frac{r}{r_s})^\alpha\rbrack^{\frac{\beta-\gamma}{\alpha}}}{dr}.
\label{scaleden}
\end{eqnarray}

Finally, by combining the above eqs. (\ref{DC14}-\ref{scaleden})  we obtain a density profile as a function of three parameters $r_{s}$, $M_{halo}$ and $M_D$, which we use in order to define the RC curve for the DC14 DM profile.

Despite the complexity of the proposed scheme, it is worth  to test such DM density profile based on the analysis of hydro-dynamically simulated galaxies \citep[see][]{dicintio14} drawn from the MaGICC project \citep{brook12,stinson13}.

\subsection{Results}

 \begin{figure*}
 \begin{centering}
\includegraphics[angle=0,height=7truecm,width=11truecm]{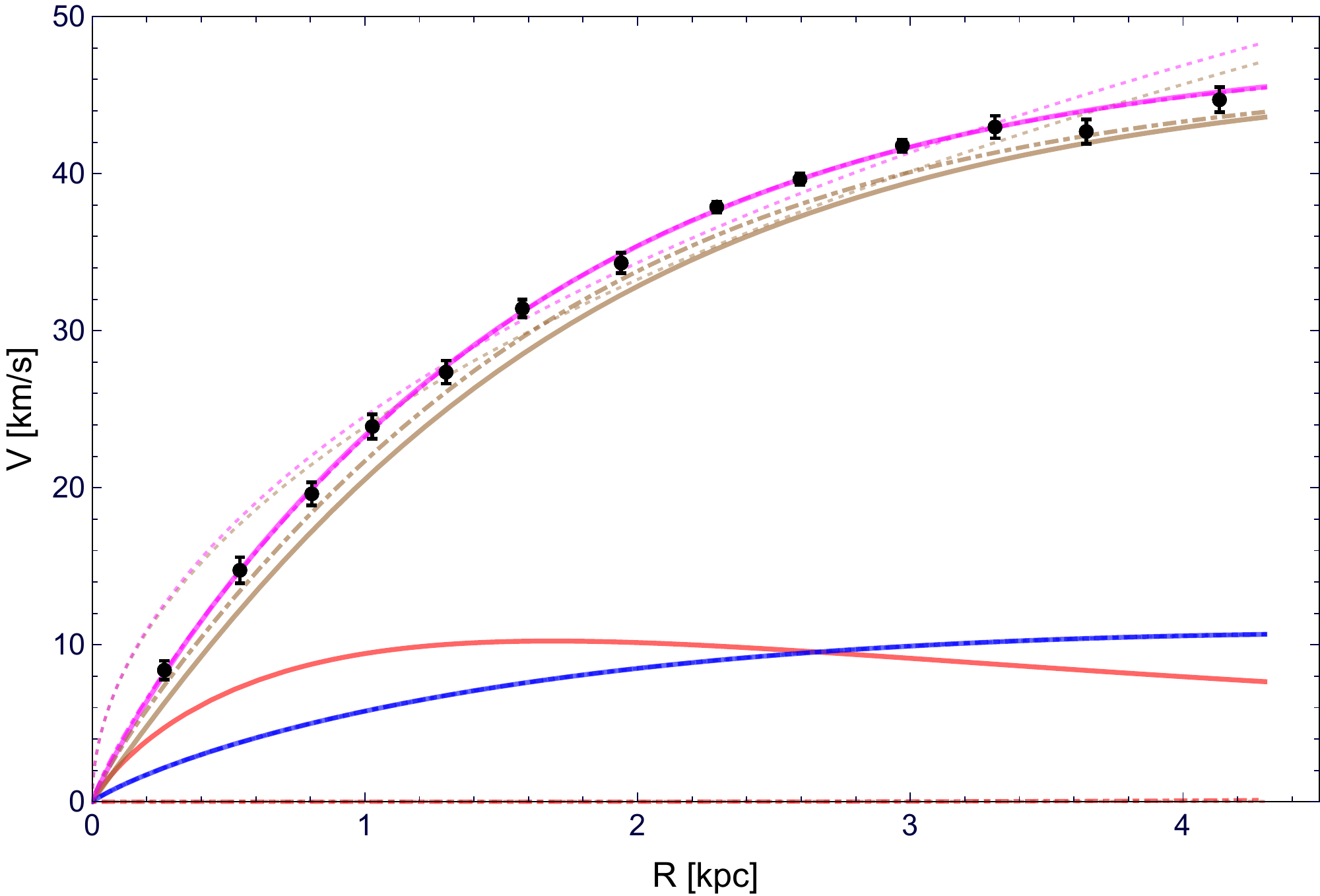}
\caption{The synthetic RC (filled circles with uncertainties) and URC with its separate dark/luminous contributions (red line: disc; blue line: gas; brown line: halo; pink line: the sum of all components)
 in case of three DM profiles: the Burkert DM profile (solid lines), NFW profile (dashed lines) and DC14 profile (dot-dashed line).}
\label{fig5}	
\end{centering}
\end{figure*}

In Fig. \ref{fig5} we show the results of the mass modelling of the fiducial RC  by means  of  the dwarf disc universal rotation curve ("dd"URC) model: 
an exponential Freeman disc, a gaseous disc plus a Burkert halo profile. This result is very successful (see solid lines of Fig. \ref{fig5}) with $\chi^2_{red} <1$. 
The best-fit parameters of the fiducial RC are: 

\begin{eqnarray}
log\langle \rho_0\rangle &=& 7.55 \pm 0.04  \quad M_{\odot}\,kpc^{-3};\nonumber\\
\langle r_c\rangle&=& 2.30 \pm 0.13 \quad kpc;\nonumber\\
log\langle M_{D}\rangle &=& 7.71 \pm 0.15 \quad M_{\odot}.
\label{fitparbur}
\end{eqnarray}

The resulting virial mass  is $ \langle M_{vir} \rangle=(1.38 \pm 0.05) \times 10^{10} M_{\odot}$. 

It is worth to recall that the coadded  double normalized RC of dds   (Table \ref{tbl:2} columns  3-5)  would be identically well fitted and the relative structure parameters can easily be obtained via the transformation laws in eq. (\ref{scaleopt}).

NFW profile fails to reproduce the synthetic RC (see dotted lines of Fig. \ref{fig5}), the reduced chi-square is $\approx 12$ and the best-fit parameters

\begin{eqnarray}
log\langle M_{vir}\rangle &=& 11.68 \pm 0.87 \quad M_{\odot};\nonumber\\
\langle c\rangle &=& 4.73 \pm 3.19; \nonumber\\
log\langle M_{D}\rangle&=& 2.50_{-2.50}^{+?} \quad M_{\odot}.\nonumber
\label{fitparnfw}
\end{eqnarray}

lead to totally unrealistic estimates of the stellar disc and halo masses.

 The DC14 profile shows the same good quality fit (see dot-dashed lines of  Fig. \ref{fig5}) as the URC profile with $\chi^2_{red}<1$ and quite similar values of  the structural parameters.  
 
\begin{eqnarray}
log\langle M_{vir}\rangle &=& 10.30 \pm 0.02 \quad M_{\odot};\nonumber\\
\langle r_s\rangle &=& 2.05 \pm 0.13; \nonumber\\
log\langle M_{D}\rangle&=& 7.30 \pm 0.14 \quad M_{\odot}.\nonumber
\label{fitpardc14}
\end{eqnarray}

Then, in spite of the fact that galaxies in our sample vary by $\sim 6$ magnitudes in the $I$ band, we obtain a universal function of the normalized galactocentric radius, similar to that set up in PSS, that is able to fit well the double normalized coadded RCs of galaxies, when extrapolated to our much lower masses.

To summarise, we have worked out the "dd"URC,  i.e. an analytical model  for the  dwarf discs coadded curve, that represents the RC of  dd galaxies. This function is given by eqs. \ref{circvel},\ref{freemandisc},\ref{burmass} and by eq. \ref{fitparbur}.  Let us stress here that the "dd"URC can be considered as the 12th bin of the URC.

\section{Denormalisation of the "dd"URC mass model}

   \begin{figure*}
 \begin{centering}
\includegraphics[angle=0,height=7truecm,width=11truecm]{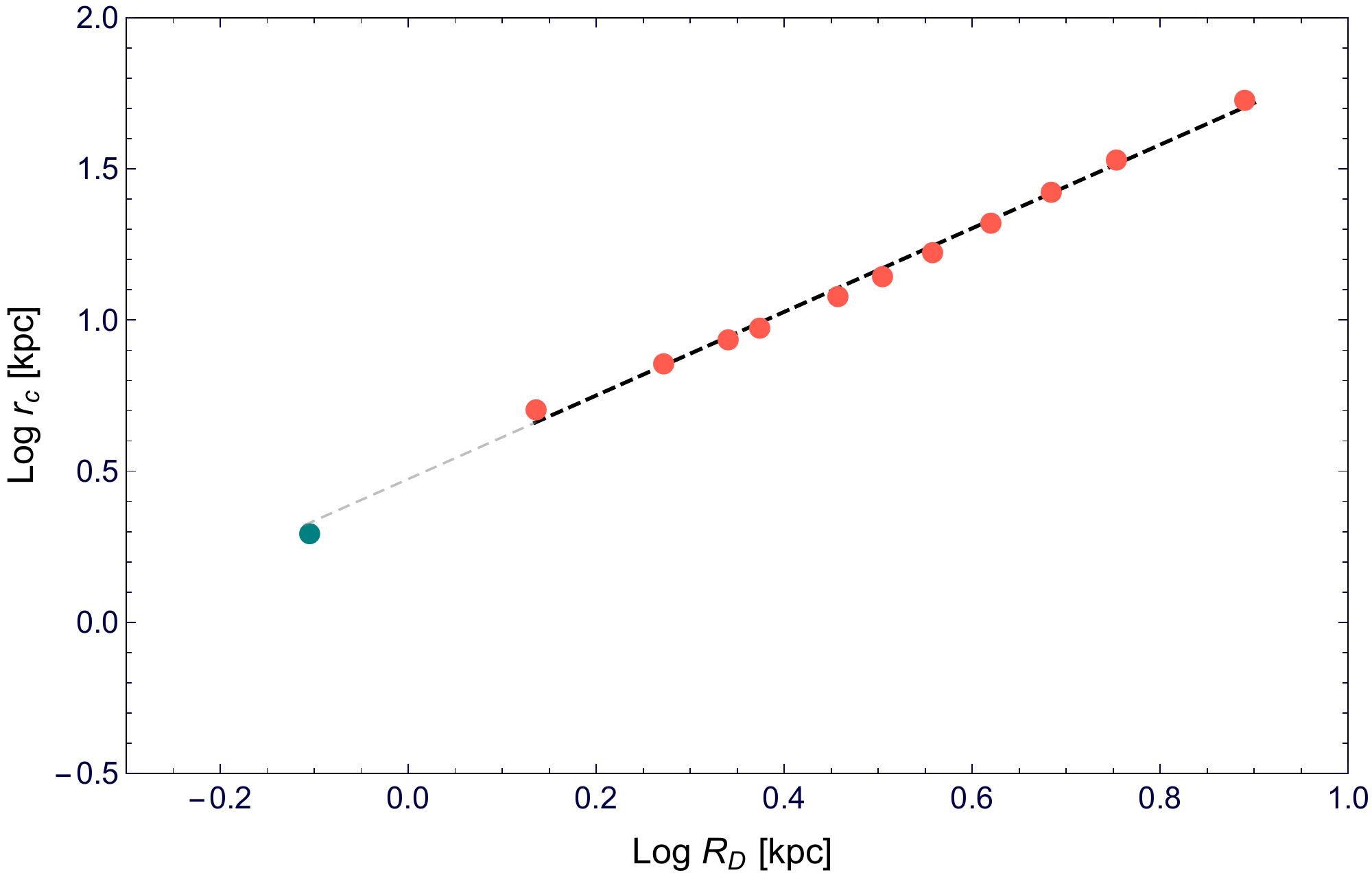}
\caption{The core radius versus disc scalelength. Red circles represent the values of the URC of normal spirals and green circle
 represents the best fit values found in the previous section. Black dashed line is a linear fit to the data of the URC of normal spirals and the grey dashed line is the extrapolation of the linear fit to the dwarf discs regime.}
\label{fit_fig4}	
\end{centering}
\end{figure*}

 \begin{table*}
\caption{Sample of dd galaxies. Columns: \textbf{(1)} galaxy name;  \textbf{(2)} the stellar disc mass; \textbf{(3)} 
the stellar disc mass using K-band luminosities; \textbf{(4)} the gas mass; \textbf{(5)}the gas mass listed in Karachentsev et al. (2013); \textbf{(6)} 
the core radius; \textbf{(7)} the central density; \textbf{(8)} the halo mass; \textbf{(9)} compactness of the stellar disc.}
\label{tbl:3}
\centering
\begin{tabular}{ l c c c c c c c c c c c c} 

	\hline
	Name  &  $M_D$  & $M_D(K_S) $ &  $M_{HI}$  & $M_{HI} (K13)$& $ r_c$ & $log(\rho_0)$& $M_h$ &C \\ ---  &$\times 10^7$  & $\times 10^7$ & $\times10^7$ &  $\times 10^7$ & ---& ---&$\times 10^9$& --- \\ ---  &$ M_{\odot}$ & $ M_{\odot} $ & $  M_{\odot} $&  $M_{\odot}$& kpc& $g/cm^3$& $M_{\odot}$&--- \\
	(1) & (2) & (3) & (4) & (5) & (6) & (7) & (8) & (9)  \\
     \hline\\    

      UGC1281 &9.8  &  19.9&   38.7 & 22.1 & 2.93&-23.6&32.4&1.1\\   
      UGC1501 & 11.3 & 23.9   &  44.3 &  38.4 & 4.32&-23.9&40.2&0.88\\
      UGC5427 &  3.74 &   8.28 &  14.7 &  3.93& 0.76&-22.5&8.89&1.86\\
%     DDO99 &   &$H_{\alpha} - 1$  &   0.27 &  c& 12.5\\
%     UGC4305 &  & $H_{I} - 2$, $H_{I} - 3$  &  1.11&  d& 34.5\\
      UGC7559 &4.19& 7.21  &  16.5 &  13.9& 2.46&-23.8&11.9&0.84\\
      UGC8837 & 12.0 &  24.4  &   47.2 &  29.8 & 5.40&-24.1&44.6&0.77\\
      UGC7047 &   2.65& 11.4  &  10.4 &  15.3 & 1.34&-23.3&6.53&1.05\\
      UGC5272 &  13.2& 6.58  &   53.0 &  23.1 & 4.14&-23.8&48.1&0.98\\
      DDO52 & 16.0 &  14.7  &   62.9 &  27.8 & 4.24&-23.8&60.1&1.05\\
      DDO101 &11.1& 49.9   &   43.7 &  16.0 & 2.71&-23.4&36.8&1.22\\
      DDO154 &  3.70 & 2.33  &   14.6 &  25.3 & 1.98&-23.6&10.0&0.93\\
      DDO168 &  10.2& 8.28  &   40.2 &  29.8 & 2.28&-23.3&32.5&1.33\\
      Haro29 &   1.02& 3.96 &  4.02 &  7.65 & 0.51&-22.6&2.02&1.37\\
      Haro36 & 10.6& 13.8  &  41.7& 14.9 & 2.84&-23.5&35.1&1.16\\
      IC10 &   2.19& 17.7  &   8.61 &  13.3& 0.78&-22.8&4.94&1.43\\
      NGC2366 &13.3   & 28.1 &   52.1 &  54.2& 4.16&-23.8&48.2&0.97\\
      WLM &   2.05 &  2.94   &   8.05 &  9.0 & 1.29&-23.4&4.86&0.96\\
      UGC7603 &13.8    & 53.5    &   54.4 &  55.4 & 3.42&-23.6&49.0&1.14\\
      UGC7861 & 7.87 & 97.3 & 30.9 &  41.1& 1.51&-23.0&22.6&1.59\\
      NGC1560&  11.8&  31.5   &   46.5 &  142.5& 3.37&-23.7&40.9&1.08\\
      DDO125 &   0.49& 7.55  &   1.91 & 4.02& 1.1&-23.8&0.93&0.56\\
  %    DDO190 &  0.53 &  4.77  &   2.16 &  4.2 & 0.55&-23.0&0.90\\
  %    KK149 &  0.97 & 9.29  &   3.93&  7.8& 0.67&-23.0&1.80\\
      UGC5423 &2.77   & 15.4  &  10.9 &  9.2& 1.19&-23.14&6.76&1.17\\
%      UGC6456 &  & $H_{\alpha} - 1$, $H_{I} - 8$  &   0.51 &  b& 14.7\\
      UGC7866 &   1.53& 9.29   &   6.02 &  10.6& 1.27&-23.5&3.49&0.85\\
      DDO43 &   2.42& 2.44  &   9.72 & 9.42& 1.35&-23.3&5.92&1.0\\
      IC1613 &   0.74& 7.05   &   2.91 &  7.8& 1.46&-23.9&1.53&0.55\\
      UGC4483 & 0.28 &0.6  &  1.09 & 4.4& 0.29&-22.6&4.513&1.12\\
      KK246 & 2.38 & 3.96  &  9.35 & 15.6& 1.40&-23.4&5.82&0.98\\
%      UGC3476 &  & $H_{\alpha} - 1$   &   0.25 &  f& 9.0\\
      NGC6822 & 2.34 &13.1  &   9.21 & 18.8& 1.32&-23.3&5.68&1.01\\
      UGC7916 &    7.63& 3.79  &  30.0& 35.8& 5.80&-24.4&26.3&0.59\\
      UGC5918 & 8.43 & 12.3   &   33.2 &  23.1& 3.88&-23.9&28.3&0.83\\
%      DDO133 &   &$H_{I} - 3$    &   1.71 &  b& 40.84\\
%      DDO216 &   &$H_{I} - 3$   &   0.37 & b& 13.25\\
      AndIV &   1.68&0.77  &   6.62 &  27.8 & 1.06&-23.2&3.81&1.02\\
%       2.0 &   86.2   &   79.0 &  7.0& ---\\
	UGC7232 & 0.99 &  3.91  &   4.0 &  3.84 & 0.34&-22.2&1.88&1.77\\
	DDO133 & 5.53&10.4 &21.7 & 21.1 &2.55&-23.7& 16.4&0.93\\
	UGC8508 & 0.62 &2.43 &2.48 & 2.65 &0.50&-22.8& 1.15&1.10\\
	UGC2455 & 8.03 &122.5 &31.5 & 87.9 &3.21&-23.8& 26.0&0.93\\
	NGC3741 & 0.33 &1.44&1.31 & 10.1 &0.27&-22.4 &0.55&1.31\\
	UGC11583 & 10.9 &5.73 &42.9 & 24.8 &3.67&-23.8& 37.8&0.97\\

                  \hline
\end{tabular}
\end{table*}

In this section we will construct a URC for the dwarf disc galaxies in the physical units that will cope with the diversity of RCs evident in Fig. (\ref{fig2}). In  spirals (see PSS) we can easily go back from  a double normalized URC $V(R/R_{opt})/V_{opt}$ to an RC expressed in  physical units  $V(R/kpc,M_I)  km/s$,  where $R_{opt}$, $V_{opt}$ and $M_I$  are altogether  well correlated. This is not the case for dds where another quantity, the compactness, enters in the above 3-quantity link.

Let us first remind that  in every radial bin the residuals do not correlate with the optical velocity of the corresponding galaxy (see Section 3). This implies that the dds structural parameters of the dark and luminous matter have  a negligible direct dependence on luminosity/optical velocity different from that inherent to the  two normalizations we apply  to the individual RCs.

Moreover, given the  very small  intrinsic scatter of the fiducial double normalized coadded RC and the extremely  good fit of the "dd"URC to it,  we can write

\begin{eqnarray}
\frac{M_{D,HI}}{V^2_{opt} R_{opt}} &=&\frac{\langle M_{D,HI}\rangle}{\langle V^2_{opt}\rangle \langle R_{opt}\rangle}\equiv const.
\label{assump}
\end{eqnarray}

\noindent
Then, we derive in all objects a direct proportionality between the halo core radius $r_c$ and the disc scale-length $R_D$, which is  in agreement with the extrapolation of the corresponding relationship in normal spirals of much higher masses \citep{salucci2007}: $log(r_c)=0.47+1.38 log(R_D)$, see Fig.~\ref{fit_fig4}.
    
We also assume that $\frac{V^2_D(R_{opt})}{V^2_{HI}(R_{opt})}$ is constant among galaxies and  it equals to the value of  the average case $\frac{\langle {V^2_D(R_{opt})}\rangle}{\langle V^2_{HI}(R_{opt})\rangle}\simeq1.1$.

Consequently with the above assumptions, for each galaxy of the sample, we have

\begin{eqnarray}
M_{DM}(R_{opt})&=&(1-\alpha) V_{opt}^2 R_{opt} G^{-1},
\label{Mdm}
\end{eqnarray}

 \noindent
where $M_{DM}$ is the Burkert DM mass inside the optical radius $R_{opt}$ and $\alpha$ is the fraction which baryonic matter contributes to the total circular velocity:

\begin{eqnarray}
\alpha&=&\frac{\langle V_{HI}^2(R_{opt})\rangle+\langle V_{D}^2(R_{opt})\rangle}{\langle V_{tot}^2(R_{opt})\rangle}
\nonumber \\
&=&0.12\equiv const.
\label{alpha}
\end{eqnarray}

\noindent
Notice that  in some galaxies  the fractional  contribution to $V$  from the HI disc can be different from the assumed value of $\sim0.06$. However, this has no effect on our investigation. In fact, at the radii where the HI disc is more relevant that the  stellar disc,  the contribution of the DM halo becomes overwhelming \citep{evoli11}.

By simple manipulations of  eqs. (\ref{assump}-\ref{alpha}) inserting the individual  values of $R_{opt}, V_{opt}$,   we  get,  for each galaxy, the structural parameters of the dark and the luminous matter.
In Table \ref{tbl:3} we list them  along side with their uncertainties  obtained from those  of the  URC mass model    given in eq. (\ref{fitparbur}).

\subsection{HI gas mass and stellar mass}

We now check the validity of the assumptions in the previous subsection. We compare our estimated values of the defined galactic HI masses, eq. (\ref{assump}), with those given by K13 (calculated using total $H_I$ flux, for more details see K13). We find: 

\begin{eqnarray}
log M_{HIkin}&=& (-0.015 \pm 1.12)
\nonumber\\
&+&(1.0 \pm 0.14) log M_{HIK13}
\label{MGcor}
\end{eqnarray}

 \noindent
with an rms of 0.3 dex.  The value  of the slope and  the  small  rms.,  despite  the presence of some  outliers  most probably originating from the large range in  luminosities and morphologies of our sample, suggest that  $M_{HIkin}$ are good proxies of $M_{HIK13}$. Therefore, adopting them does not influence any result of this paper.

We also compare the kinematical derivation of the stellar disc masses for the objects in our sample with those  obtained for the same objects from $K_S$
band photometry (provided by K13). Following  \citet{bell03,mcgaugh15} we adopt a constant mass-to-light ratio of $M/L_K=0.6 \times M_{\odot}/L_{\odot} $ and we report them in Table \ref{tbl:3} as $M_D(K_S)$. We find a good correlation between the two estimates:

\begin{eqnarray}
log M_{Dkin}&=& (2.49 \pm 1.0)
\nonumber\\
&+&(0.64 \pm 0.12) log M_{DK_S}
\label{MDcor}
\end{eqnarray}

 \noindent
\noindent
with an rms of 0.4 dex. The two estimates are therefore  mutually consistent  especially by considering that the kinematical estimate has an uncertainty of $0.3$ dex \citep[see][]{salucci08}. Let us also notice that in dds the conversion between luminosity and stellar masses is subject to a similarly large systematical uncertainty, especially in actively star-forming galaxies like those present in our sample. 

 These results, therefore, support well the scheme used in this paper to deal with the luminous components of dds.

Furthermore, we compare our results with \citet{lelli16}, where the authors analyse a sample of 176 disc galaxies and quantify for them the ratio of baryonic-to-observed velocity. We have, that this ratio, calculated at 2.2 disc scale lengths, is $\sim 0.4$. The latter is consistent with values of \citet{lelli16}, established for a sample of dwarf disc galaxies. Moreover, we found that the value of gas fraction ($f_{gas}\equiv \frac{M_{HI}}{M_{bar}}\sim 0.8$) in our sample is also consistent with the value estimated by \citet{lelli16}, where the authors show that low-luminosity end disc galaxies are extremely gas dominated with $f_{gas}\simeq0.8-1.0$.

%\textbf{Notes.} 

 \begin{figure*}
 \begin{centering}
\includegraphics[angle=0,height=6truecm,width=11truecm]{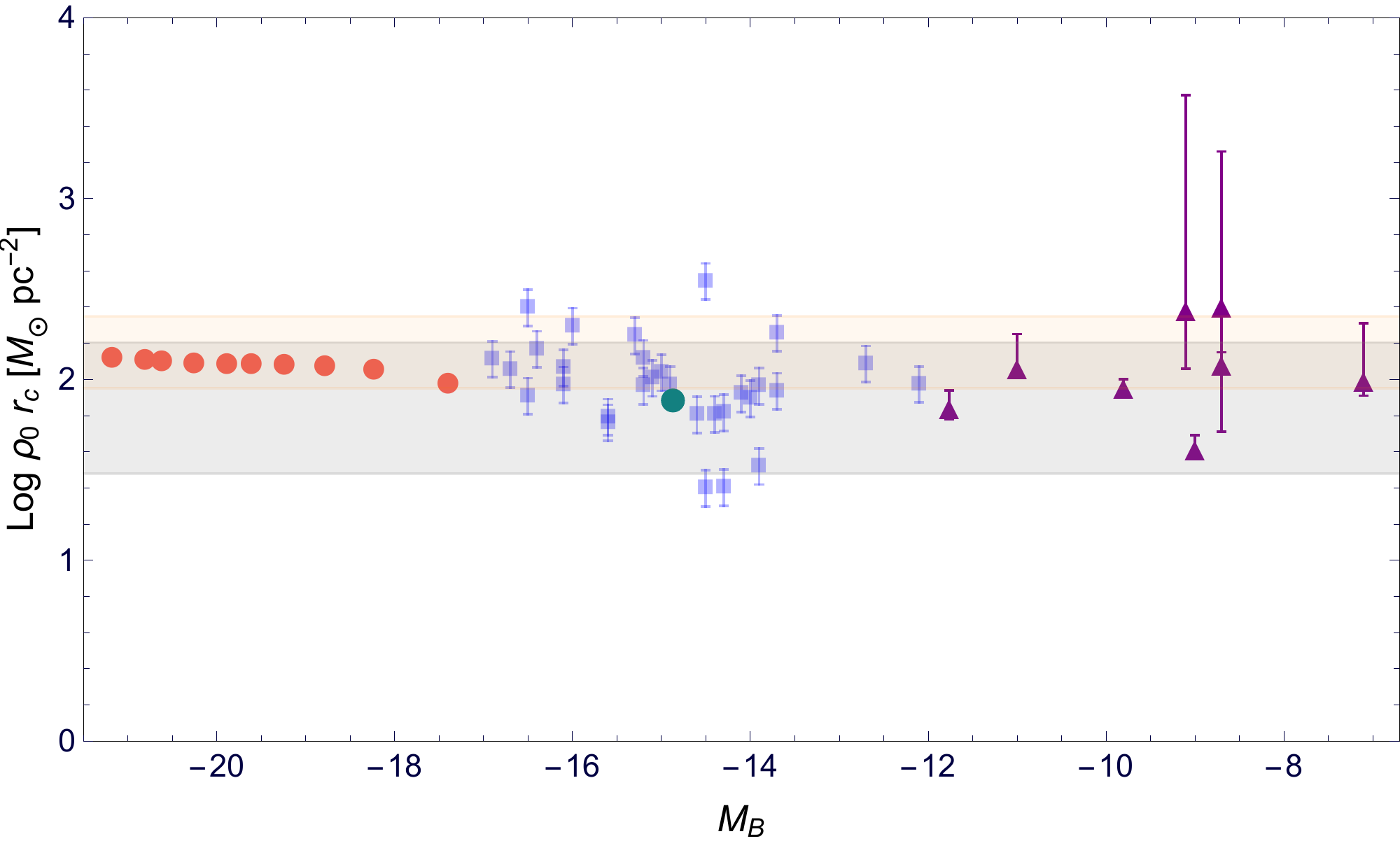}
\caption{$\rho_0r_c$ in units of $M_{\odot} pc^{-2}$ as a function of a galaxy magnitude for different galaxies and Hubble types.
 The data are: the \citet{salucci12} the URC of normal spiral galaxies (red circles); scaling relation from \citet{donato09} (orange shadowed area);
  Milky Way dSphs (purple triangles) \citet{salucci12}; dwarf disc galaxies (blue squares-this work, green dot represents the average point), B magnitudes are taking from KK13;
empirically inferred  scaling relation: $\rho_0 r_c=75^{+85}_{-45} M_{\odot} pc^{-2}$ from \citet{burkert15}(grey shadowed area).}
\label{rhor}	
\end{centering}
\end{figure*}

\begin{figure*}
 \begin{centering}
\includegraphics[angle=0,height=7truecm,width=11truecm]{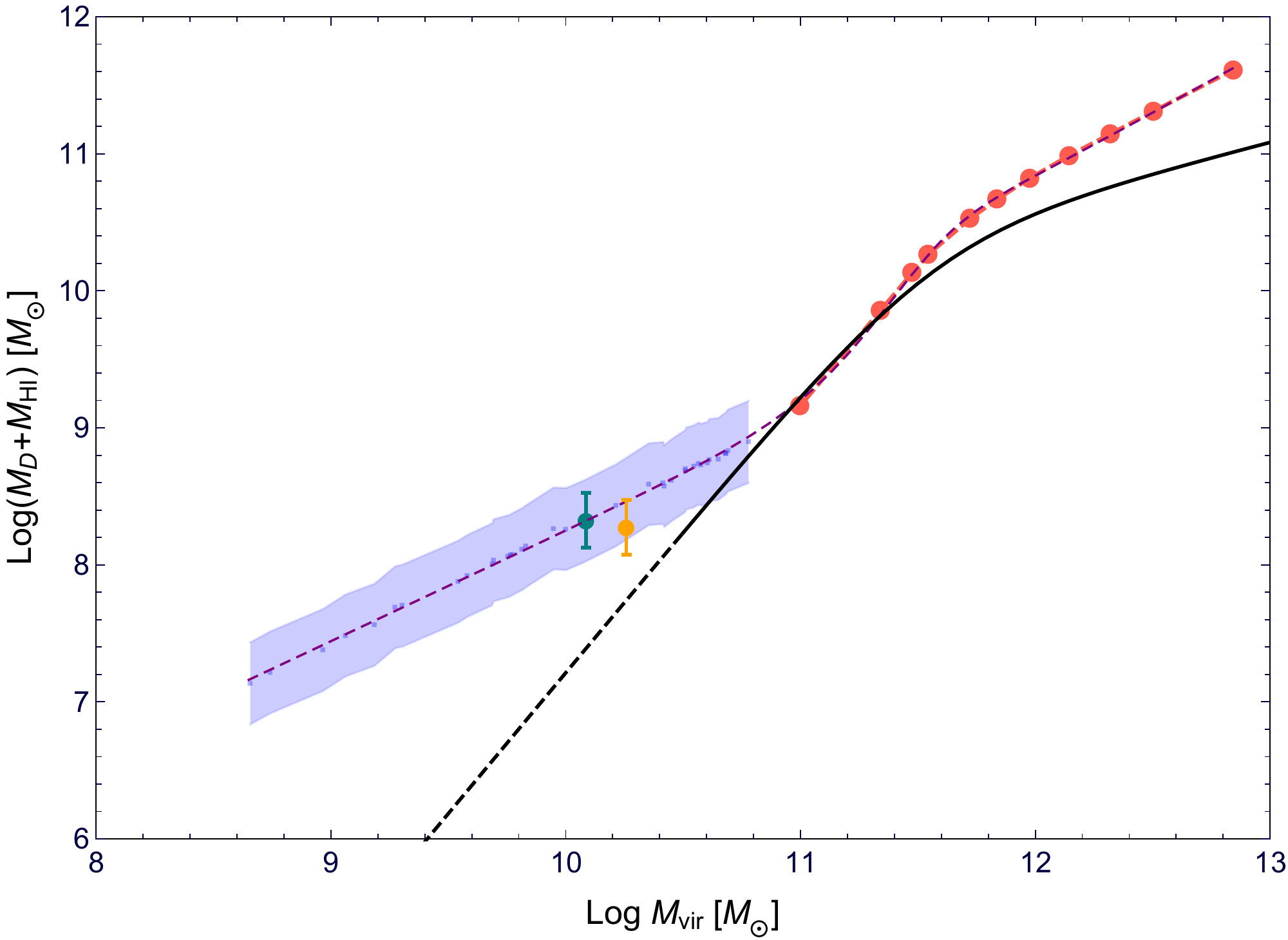}
\caption{The baryonic mass versus the virial mass for normal spirals (joined red circles) and for the dwarf discs assuming the URC model (blue shadowed area assuming 0.3
 dex scatter, the green circle with error bars represents the average point of the region). Yellow dot with error bars is the best fit value for the fiducial RC using DC14 model (see Section 4). Purple dashed line corresponds to the parameterised
  eq. (\ref{ferrero}) of the galaxy baryonic mass as a function of halo mass. The abundance matching relation from \citet{papastergis12} is shown by black solid line, the region that is extrapolated from the \citet{papastergis12} relation is dashed.
}
\label{fig6}	
\end{centering}
\end{figure*} 

\begin{figure*}
 \begin{centering}
\includegraphics[angle=0,height=7truecm,width=11truecm]{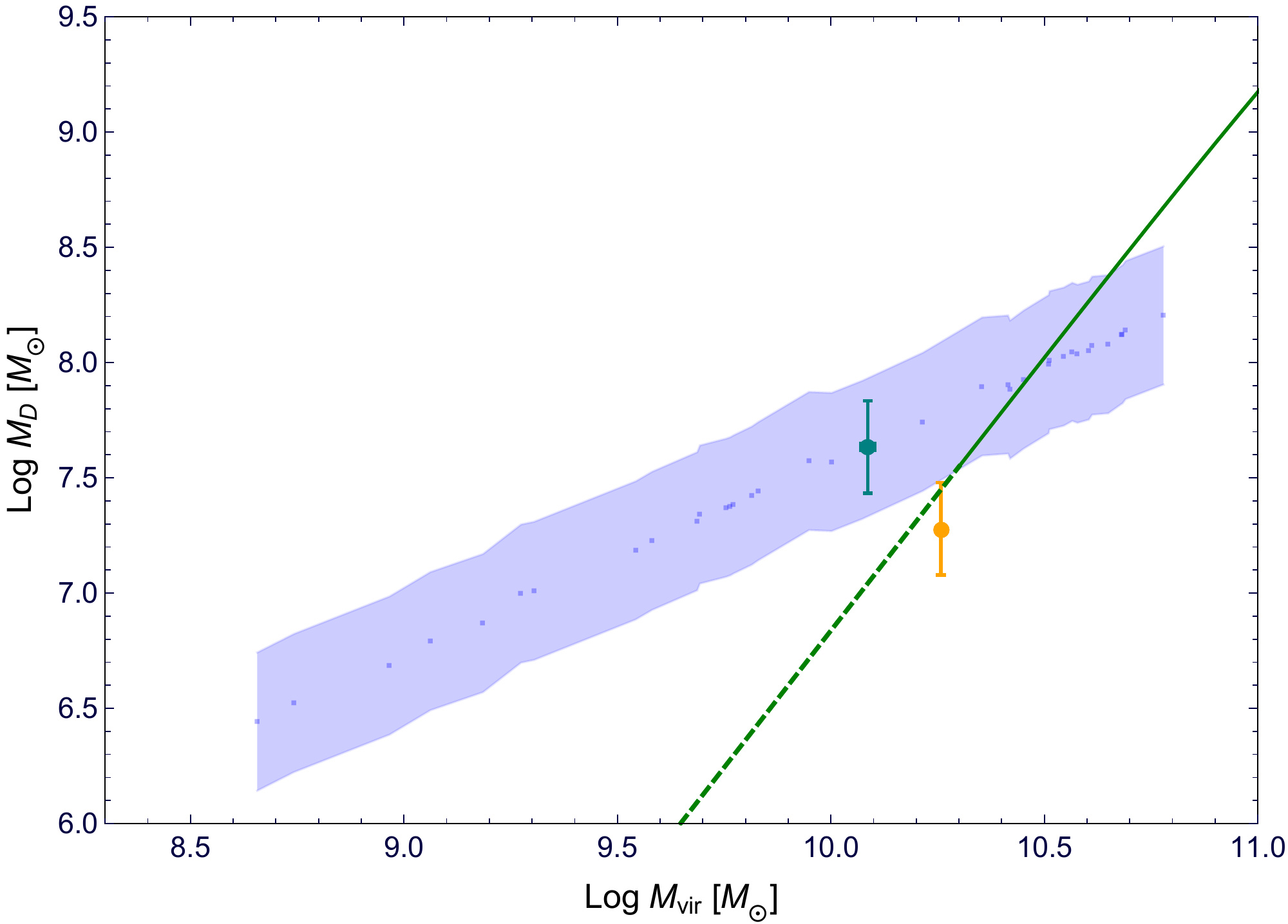}
\caption{The disc mass versus the virial mass. Blue shadowed area represents the relation for the dwarf discs assuming the URC model and taking into account 0.3 dex scatter, the green circle with error bars represents the average point of the region. Yellow dot with error bars corresponts to the best fit value for the fiducial RC using DC14 model (see Section 4). The stellar mass-to-halo mass relation from \citet{moster13} is shown by green solid line, the region that is extrapolated from the \citet{moster13} relation is dashed.
}
\label{fig6a}	
\end{centering}
\end{figure*} 
 
 \begin{figure*}
 \begin{centering}
\includegraphics[angle=0,height=6.2truecm,width=18truecm]{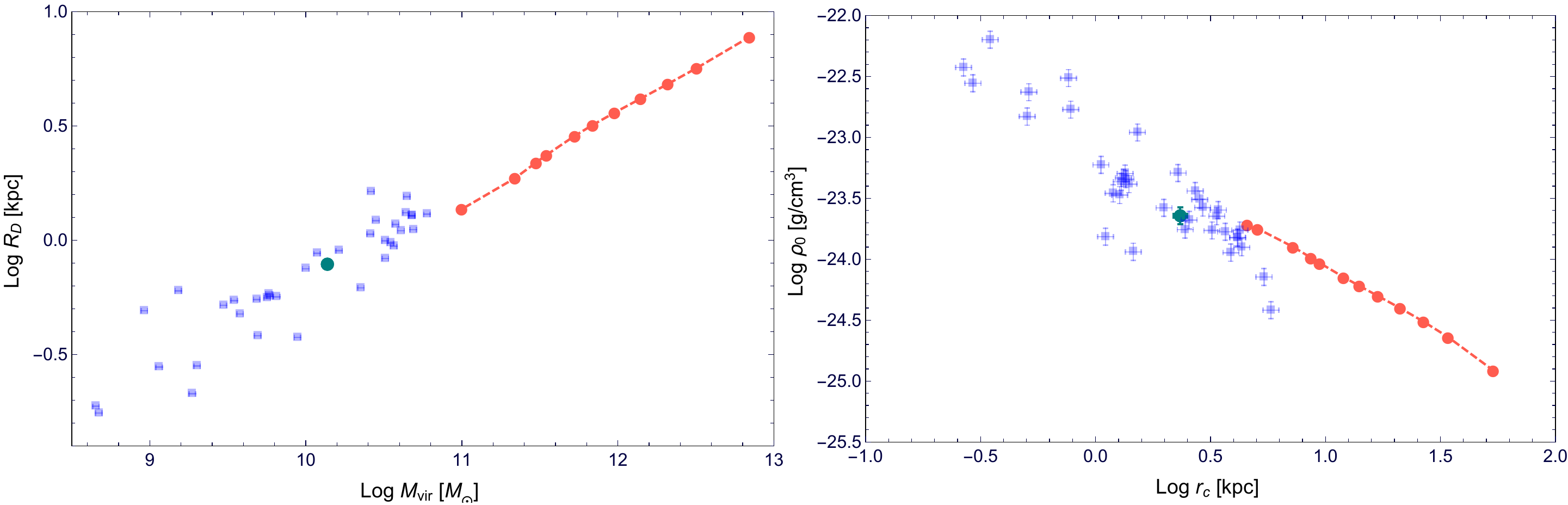}
\caption{{\it Left panel:} the disc scalelenght versus virial mass. {\it Right panel:} the central density versus core radius. Red circles represent normal spirals, blue squares with error bars correspond to dwarf discs of this work and the green circle with error bars represents the average point.}
\label{fig7}	
\end{centering}
\end{figure*}

\begin{figure*}
 \begin{centering}
\includegraphics[angle=0,height=7.truecm,width=12truecm]{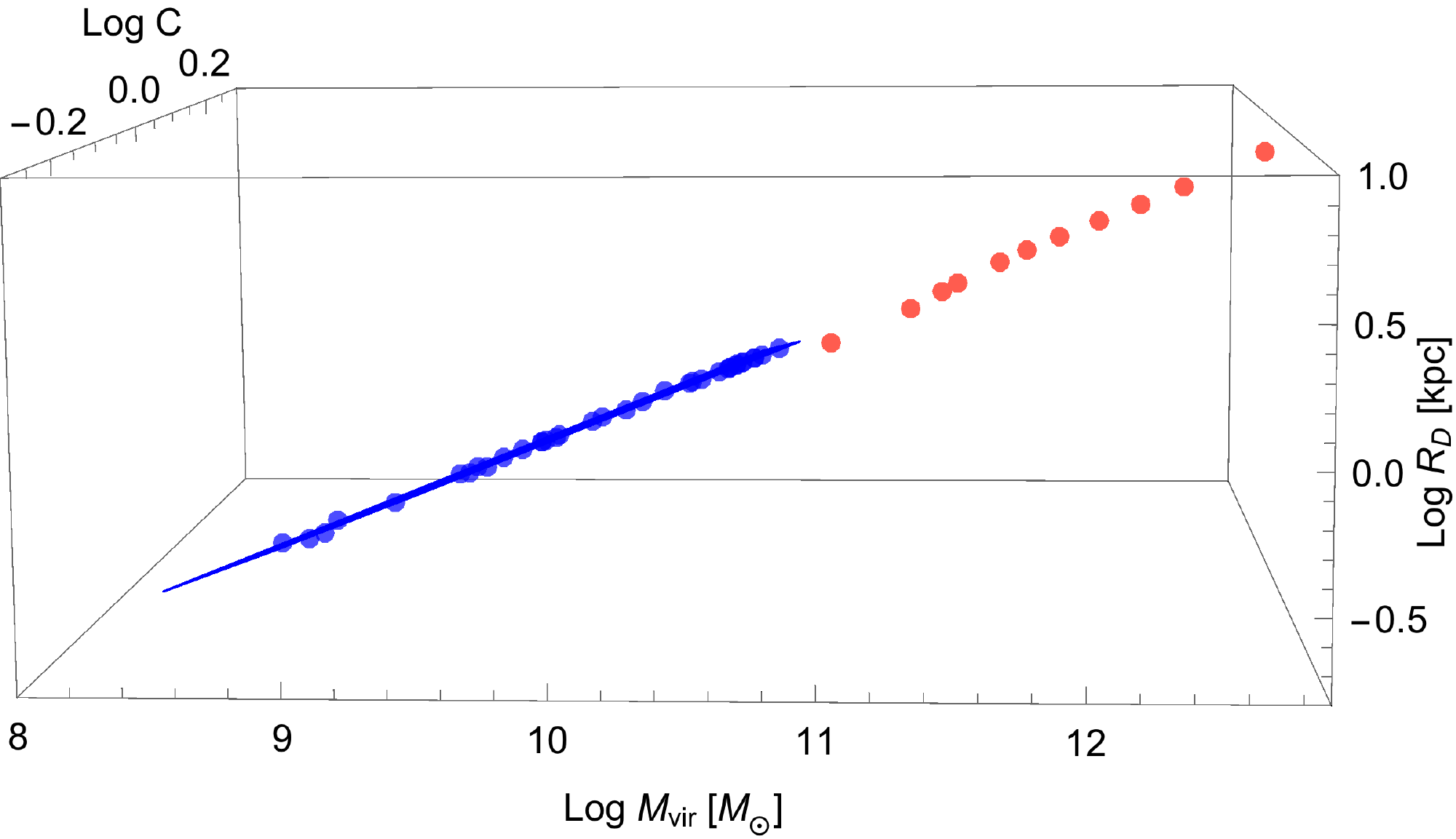}
\caption{The disc scalelenght versus virial mass and the compactness parameter C. Red circles represent normal spirals, blue squares with error bars correspond to dwarf discs of this work and blue line is the result of the fit (for details see text).}
\label{mhcrd}	
\end{centering}
\end{figure*} 

\begin{figure*}
 \begin{centering}
\includegraphics[angle=0,height=7.truecm,width=12truecm]{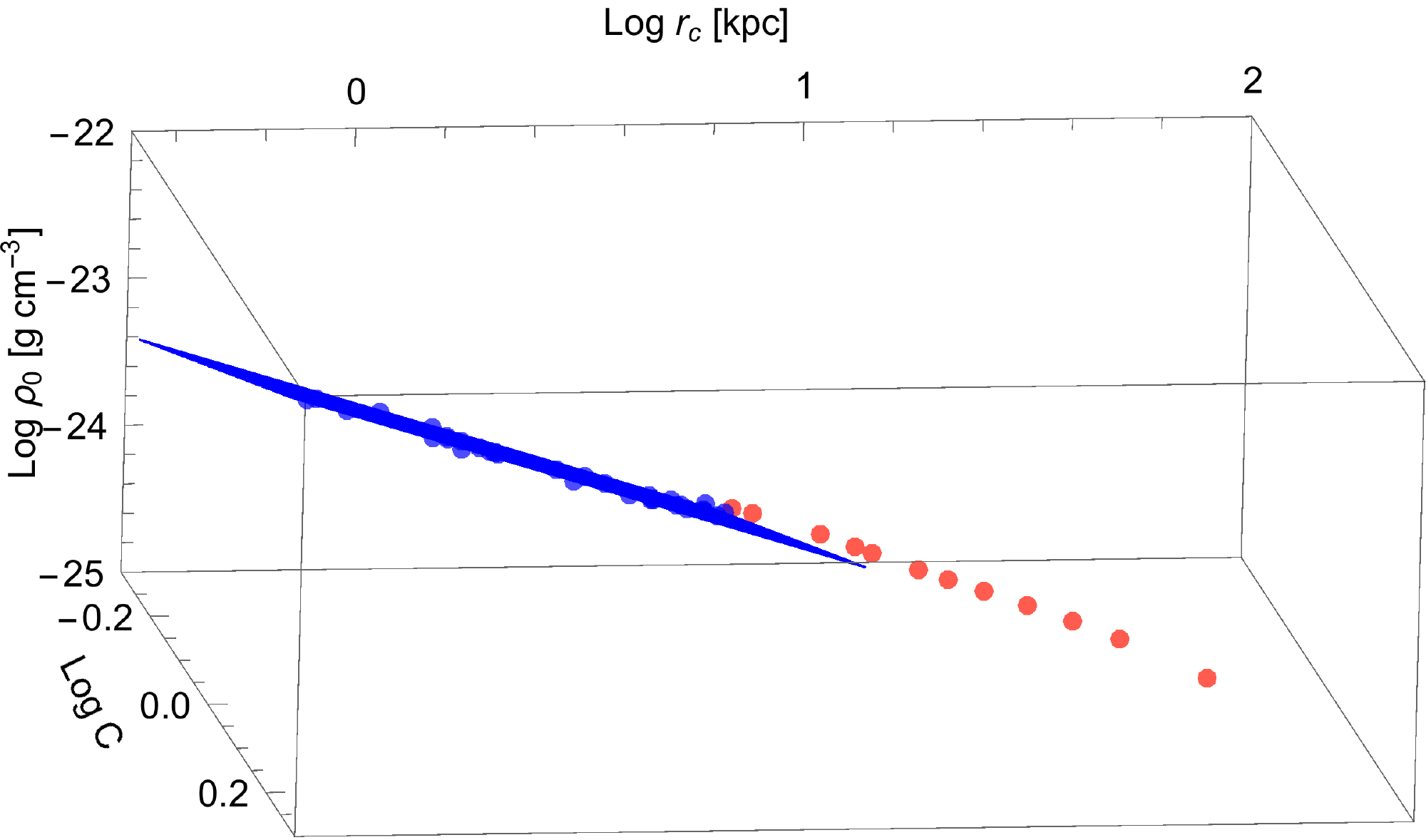}
\caption{The central density versus core radius and the compactness parameter C. The lines and symbols are as in Fig. \ref{mhcrd}.}
\label{rho0rcc}	
\end{centering}
\end{figure*} 

\subsection{The scaling relations}

Let us plot,  the central  surface density of the DM haloes of our sample, i.e. the  product of $\rho_0 r_c$,  as a function of B magnitude (see Fig. \ref{rhor}).
A constancy of this product has been found  over 18  blue magnitudes and in objects ranging from dwarf galaxies to giant galaxies \citep[e.g][]{kormendy04,gentile09,donato09,plana10,salucci12,ogiya14}.
For the case of dds, in  Fig. \ref{rhor}, one can see that most of the objects of our sample fall inside the extrapolation of \citet{donato09} relation (see the orange shadowed are of Fig. \ref{rhor}) with a scatter of about  0.3 dex of an uncertain origin.

We  now work out the relationships among the various structural properties of the dark and luminous matter of each galaxy in our sample. These will  provide us with crucial information on the relation between dark and baryonic matter as well as on the DM itself.  Obviously these relationships  are also necessary in order to establish the URC for the present sample.

We  first derive the galaxy baryonic mass versus halo virial mass relation and compare it with that of normal spiral galaxies \citep{salucci2007}, see Fig. \ref{fig6}. We take 0.3 dex as 1-$\sigma$ error in the baryonic mass (blue shadowed area). Figure \ref{fig6} highlights that galaxies of our sample, i.e. dwarf disc objects live in haloes with masses below
 $5 \times 10^{10} M_{\odot}$ and  above $4 \times 10^{8} M_{\odot}$. A similar result was found by \citet{ferrero12}, who analysed  a sample of dwarf disc galaxies either by using the individual mass modelling or the outermost values of their RCs. In Fig. \ref{fig6}, we also show the comparison of our results with the relation derived from the abundance matching method by \citet{papastergis12}. Remarkably, for  $M_{vir} \lesssim 4 \times 10^{10} M_{\odot}$, it is inconsistent with the relation found from the abundance matching method and its extrapolation (see Fig \ref{fig6}). Likewise, $M_{bar}-M_{vir}$ relation found for dwarf discs is significantly shallower than that of the low-mass spirals. The origin of this discrepancy is unclear. One possibility might be that we are facing a selection effect. This means that galaxies in our sample have, on average, more gas than that of the \citet{papastergis12} sample. We check it by excluding the gaseous mass and then comparing the stellar disc mass versus the virial mass relation of our sample with that derived from the abundance matching method \citep[we use the relation of][]{moster13}, see Fig. \ref{fig6a}. Remarkably, although there is still discrepancy between the $M_D-M_{vir}$ relation of the URC and that of the \citet{moster13}, however it shifted to the lower masses. Furthermore, let us stress that the fit resulting from the baryon-influenced DC14 profile has a lower value of the disc mass. Consequently, the derived baryonic mass against the virial mass value for the DC14 profile comes quite close to the extrapolation of the abundance matching relation of \citet{papastergis12}, see Fig. \ref{fig6}. Along with that, the derived stellar disc mass against the virial mass value for the DC14 profile agrees, within the errors, with the extrapolation of the abundance matching relation of \citet{moster13}, see  Fig.~\ref{fig6a}. However, to investigate properly this issue we should derive the transformations lows for the DC14 profile similar to that of the Burkert profile described in Section 5. The latter is beyond the scope of this paper and we are going to address this in a future work.
 
   \begin{figure*}
 \begin{centering}
\includegraphics[angle=0,height=7.truecm,width=12truecm]{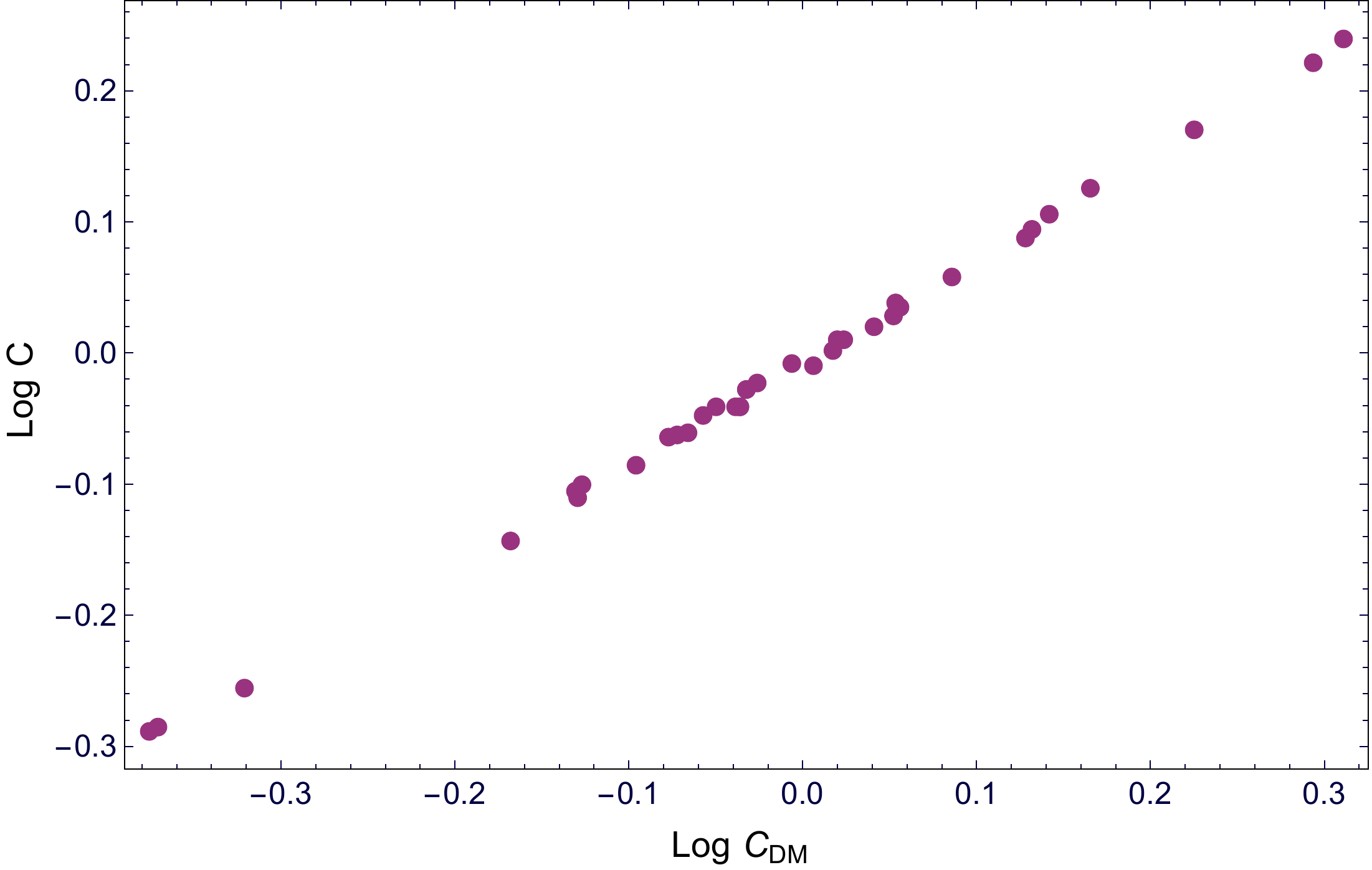}
\caption{The compactness of the stellar disc versus the compactness of the dark matter halo.}
\label{cvscdm}	
\end{centering}
\end{figure*}

   Notice, that there is also some discrepancy (irrelevant to the results of this paper) in the baryonic to halo mass relation also at higher masses. The latter is most likely due to the difference of Hubble types in the samples and in the analysis used to obtain these relations.

 Then we derive our galaxy baryonic mass versus halo virial mass relation by fitting it with the function of 7 free parameters advocated by \citet{ferrero12} :

\begin{eqnarray}
M_{bar}&=&M_{vir} \times A \left(1+\left(\frac{M_{vir}}{10^{M_1}}\right)^{-2}\right)^{\kappa}
\nonumber\\
&\times&\left(\left(\frac{M_{vir}}{10^{M_0}}\right)^{-\alpha}+\left(\frac{M_{vir}}{10^{M_0}}\right)^{\beta}\right)^{-\gamma},
\label{ferrero}
\end{eqnarray}

 \noindent
we found  $A=0.070$, $\kappa=1.85$, $M_1=11.34$, $M_0=11.58$, $\alpha=3.34$, $\beta=0.043$, $\gamma=1.05$ (purple dashed line of Fig. \ref{fig6}).

The other two relationships  which are necessary to establish the URC of dds  also in physical units  i.e. $R_D$ - $M_{vir}$ or $\rho_{0}$ -$r_c$, show a very large scatter (see Fig. \ref{fig7}) as  a consequence of the presence of dwarf disc galaxies in the sample (and in the Universe)  with almost the same stellar mass (luminosity)  but  with a different size of their stellar discs. At face value, relationships in Fig. \ref{fig7}  may lead us to exclude the existence of the URC in physical units for dwarf disc galaxies. In fact, the large scatter in Fig.~\ref{fig7} requires a new parameter to restore it.

 Therefore, we proceed and show that the universality is restored by introducing a new parameter, which we call "compactness" $C$. We define, for galaxies in the (dd) sample, the quantity $ C $ as the ratio between the value {\it predicted}  from the measured galaxy  disc mass  $M_D$ according to the simple linear regression $R_D$ vs $M_D$ of the whole sample and that of $R_D$ measured from photometry. As regard, we find:

\begin{equation}
log R_D=-3.64+0.46\hspace{2px} log M_D.
\label{rdmd}
\end{equation}

Then, we obtain the following expression of $C$,
   
\begin{equation} 
C=\frac{10^{(-3.64+0.46\hspace{2px} log M_D)}}{R_D}
\label{c}
\end{equation}
that obviously describes the  differences of the sizes of the stellar discs reduced at a same stellar mass.   
$C$ varies from 0.96 to 1.02 and its distribution in our sample is listed in Table \ref{tbl:3}.

By fitting  $log R_D$ to $log M_{vir}$ with an additional variable $log\hspace{2px}C$, we obtain an excellent fit shown in Fig. \ref{mhcrd}. The model  function being, 

\begin{equation}
log R_D=-3.99+0.38\hspace{2px} log M_{vir}-0.94\hspace{2px} log\hspace{2px}C.
\label{rdmhc}
\end{equation}

\noindent
this relation just acknowledges the existence of another player  in the stellar disc mass-size interplay.

Then we fit $log \rho_0$ to $log M_{vir}$  and $log\hspace{2px}C$:

\begin{equation}
log \rho_0=-18.26-0.51\hspace{2px} log M_{vir}+3.44\hspace{2px} log\hspace{2px}C.
\label{rhomhc}
\end{equation}

Finally, we fit $log \rho_{0}$ -$log r_c$  by adding  $log\hspace{2px}C$ as a free parameter. The result of the fit is shown on Fig. \ref{rho0rcc} and the model  function is,

\begin{equation}
log \rho_0=-23.14-0.97\hspace{2px} log r_c+2.18\hspace{2px} log\hspace{2px}C.
\label{rccrho0}
\end{equation}

It is remarkable that a basic property of the stellar discs enters to set the relationship between two DM structural quantities.  
Therefore, the scatter, which appears in dwarf discs when we try to relate the local properties of either baryonic or DM can be eliminated by using an additional
 parameter $C$. Let us note here, that very few galaxies of PSS's least luminous bin have structural properties that overlap with those of the galaxies in our sample (see, e.g., Fig. \ref{fig2}).
In a future work we will investigate the exact  details of the onset of the $C$ compactness  regime. This is clearly necessary  for constructing the URC and, on the theoretical side,  the  need  of an  additional parameter in the mass model of the late-type galaxies has to have important implications. At the same time, it will be also interesting to investigate  whether, in  normal spirals of PSS sample,  the $C$ compactness plays any role in defining the URC. Notice, however,  that we already know,  in view of the small scatter of the Radial Tully relation \citep{yegorova07}, that the spirals RCs  have a small dependence from  another parameter beyond the luminosity/mass.

Then, analogously to the compactness of the stellar disc, we define  $C_{DM}$ as the compactness of the DM halo. This quantity is the ratio, galaxy by galaxy, between the DM core radius $r_c$ (see column 6 of Table \ref{tbl:3}) and the predicted value that we obtain from the simple linear regression between  $r_c$ and  $M_{vir}$, which reads:

\begin{equation}
log r_c=-5.08+0.53\hspace{2px} log M_{vir}.
\label{rcmvir}
\end{equation}

Then, we have:

\begin{equation}
C_{DM}=\frac{10^{-5.08+0.53\hspace{2px} log M_{vir}}}{r_c}.
\label{cdm}
\end{equation}

We find that the compactness of the stellar disc is closely related to the compactness of the DM halo, see Fig. \ref{cvscdm}. Consequently, the  DM and  the stars distributions follow each other very closely. This is extremely remarkable: it may indicate a non standard  nature of the dark matter  or the fact that baryonic feedbacks easy the cusp core problem in a WIMP scenario \citep[see, e.g.,][]{teyssier13,elbadry16,dutton16,dicintio16}.

 Finally,  by using eqs. (\ref{freemandisc},\ref{freemandiscgas},\ref{burmass},\ref{ferrero},\ref{c}-\ref{rccrho0})  we derive $V_{``dd"URC} (R,M_D,R_D,C)$  the  universal function that describes  the  dwarf disc  RCs in physical units. Differently, from galaxies of higher masses it has 3 parameters, disc mass $M_D$, disc scale length $R_D$ and concentration $C$, to account for the diversity of the mass distribution of these galaxies.

%\vspace{100px}

\section{Summary and Conclusions}

We have compiled literature data for a sample of dd galaxies in the local volume ($\lesssim 11 Mpc$) with HI and $H_{\alpha}$ RCs. Then for these galaxies we establish the corresponding URC in normalized and physical units and investigate the related dark and luminous matter properties, not yet studied statistically in these objects. Our sample spans $\sim 2$ decades ($\sim 10^6-3 \times 10^8 L_{\odot}$)  in  luminosity, which coincides with the faint end of the luminosity function of disc galaxies. In magnitude extension is as large as the whole range of normal spirals usually investigated in terms of URC. For example, the galaxies in the sample are up to $\sim4$ magnitudes fainter than the lowest limit in the PSS sample.

We find that, the large variations of our sample in luminosity and morphologies require double normalization. Notably, after this noralization we have that all RCs in double normalized units are alike. This implies that the structural parameters 
of the dark and luminous matter of these galaxies do not have any explicit dependence on luminosity except those coming from the normalizing process. Additionally, the good agreement of our coadded RC with that of the first PSS's luminosity bin indicates that in such small galaxies the mass structure is already dominated by a dark halo with a density core as big as a stellar disc.

Then by applying to the double normalized rotation curve the standard $\chi^2$ mass modelling, we tested three DM density profiles. Wherein the NFW profile fails to reproduce the coadded curve, while the Burkert and DC14 profiles show excellent quality fits with $\chi_{red}^2<1.$ This result points towards the cored DM distribution in dwarf disc galaxies. The same conclusion was drawn in the papers on Things and Little Things samples \citep[see, e.g.,][]{oh11,oh15}, where the authors found for their dwarfs much shallower inner logarithmic DM density slopes than those predicted by DM-only ($\Lambda$)CDM simulations. The present analysis has the advantages of bigger statistics, but above all, is immune from systematics that can affect the mass modelling of individual galaxies.

We also defined, galaxy by galaxy, the values of the dark and luminous matter structural parameters. Surprisingly, a new actor enters the scene of the distribution of matter in galaxies, the compactness of the stellar component, which allows us to establish the URC in these low-mass galaxies. However, in order to understand better the role of this compactness it is required to investigate galaxies in the transition regime  which appears at about $V(R_{opt})\simeq60 km/s $.

As a consequence of the derived  mass distributions, there is no evidence for the sharp decline in the baryonic to halo mass relation. Similar result, for dwarf galaxies in the field, was found by
\citet{ferrero12}. Nevertheless, notice that in  DC14  case the estimated baryonic mass is slightly lower than that of the URC mass model, which brings it closer to the abundance matching relation inferred from e.g. \citet{papastergis12}. Furthermore, since the fit resulting from the baryon-influenced DC14 profile has a lower value of the disc mass, it agrees, within the errors, with the extrapolation of the $M_D-M_{vir}$ relation derived from the abundance matching by \citet{moster13}. Let us also recall, that the \citet{dicintio14} model has been already tested  against observations in works by \citet{katz16} and \citet{pace16}. Although both groups use similar methods, the drawn conclusions are different \citep[see also][]{read16}. Therefore, the consistency level between observations and the  ($\Lambda$)CDM model of galaxy formation, specifically the abundance matching technique deserves further investigation.

At the same time, the S-shape of $M_{vir}-M_{bar}$ relation may be interpreted as different physical mechanism occurring along the mass sequence of disc galaxies. Theoretically,  it has been shown that the energetics of star formation differs among different galaxies with a characteristic dependence on the halo-to-stellar mass ratio
\citep{dicintio14,chan15} and possibly also on star formation history \citep{onorbe15}.

 We remark that we found that the  DM and  the stellar distribution follow each other very closely out to the level for which, in log-log frame, the compactness of the stellar disc is proportional to that of the DM halo. We believe that here we are touching a crucial aspect in the DM issue, whose investigation, however, much exceed the scope of this paper.

Finally we would like to stress that the results of this work \citep[and of the previous works, see, e.g., ][]{donato09,gentile09} indicate that the DM around galaxies should be considered, rather than the final product of the cosmological evolution of the massive components of the Universe, galaxies, today, but as the direct manifestation of one of its  most extraordinary mysteries.
%we recall that the study of the "dd"URC has lead to the main properties of dark matter in dwarf discs and it will be integrated by future analyses of suitable  individual RCs, once they will be available in a sufficient number. 

\section*{Acknowledgements}

We thank Gianfranco Gentile, Federico Lelli, Alexei Moiseev, Se-Heon Oh and Rob Swaters for providing their data in electronic form. We would like to acknowledge Luigi Danese, Andrea Lapi and Nicola Turini  for valuable discussions. The authors grateful to  the anonymous referees for useful comments and suggestions. We thank Brigitte Grein{\"o}cker for language corrections. E.~K. acknowledges the hospitality of ICTP--SAIFR through the FAPESP process 2011/11973-4, for the final stages of this work.

\clearpage

\bibliographystyle{mnras} % style aa.bst
\bibliography{paper} % your references Yourfile.bib

\appendix{
\renewcommand\thefigure{\thesection.\arabic{figure}}    
\section{sample of rotation curves}

In Fig \ref{figa1} we show the rotation curves of all observed galaxies used in our analyses, i.e. the same galaxies as appear in Table~\ref{tbl:1}. We note that rotation curves of UGC1501,
 UGC5427, UGC8837, UGC5272, IC10, KK149 and UGC3476a are not extended to 3.2 $R_D$ (the vertical dashed grey line of Fig. \ref{figa1} indicates the position of 3.2 $R_D$ for each galaxy),
  therefore, in order to know the value of the circular velocity at these radii we made extrapolations.

\setcounter{figure}{1}

 \begin{figure*}
\centering
\label{figa1}
\includegraphics[angle=0,height=18truecm,width=13truecm]{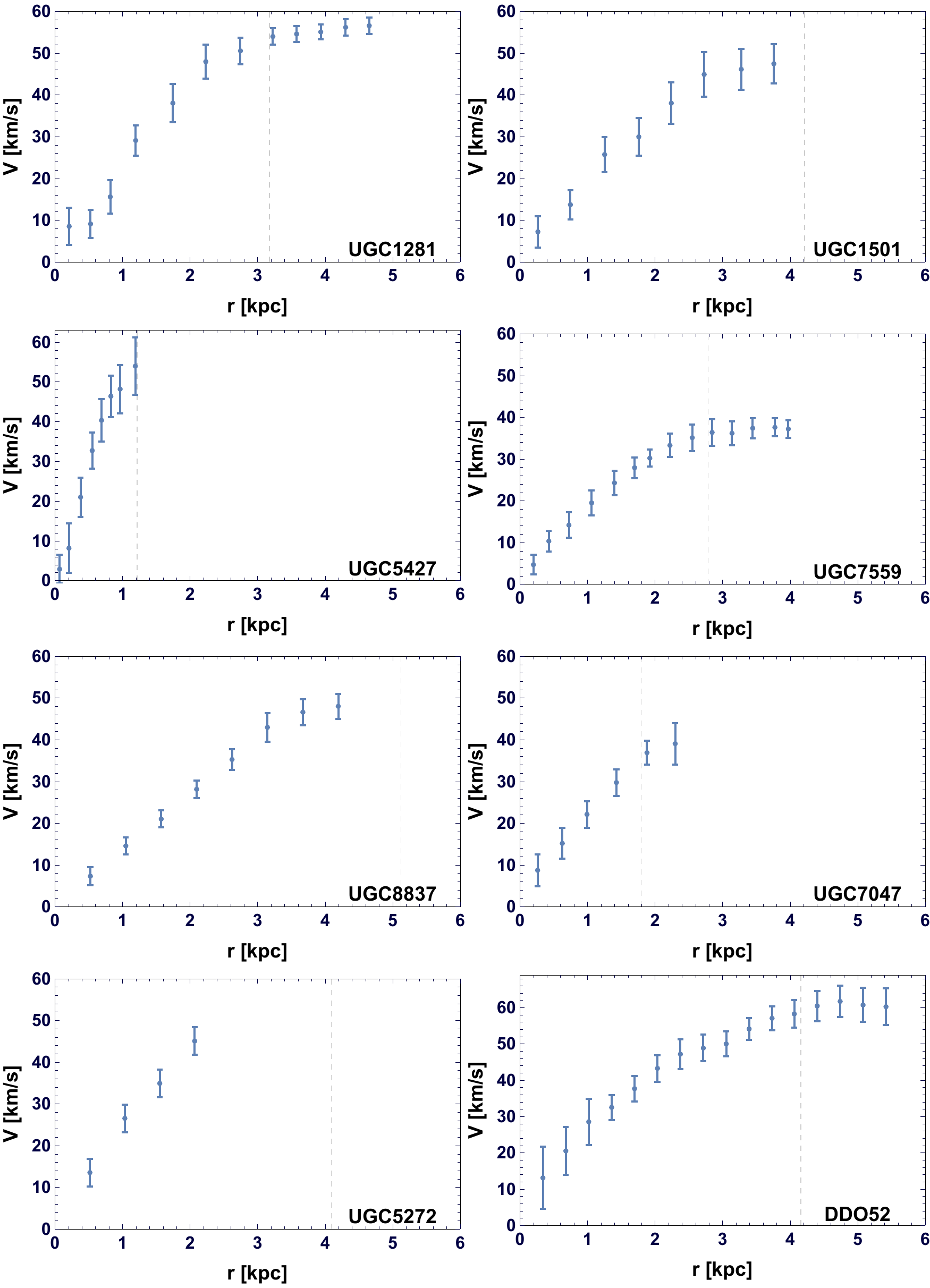}

\caption{Individual RCs. Here the $R_{opt}$ are indicated by dashed vertical lines.}
 
\end{figure*}

 \begin{figure*}
 %\ContinuedFloat
\centering
\includegraphics[angle=0,height=18truecm,width=13truecm]{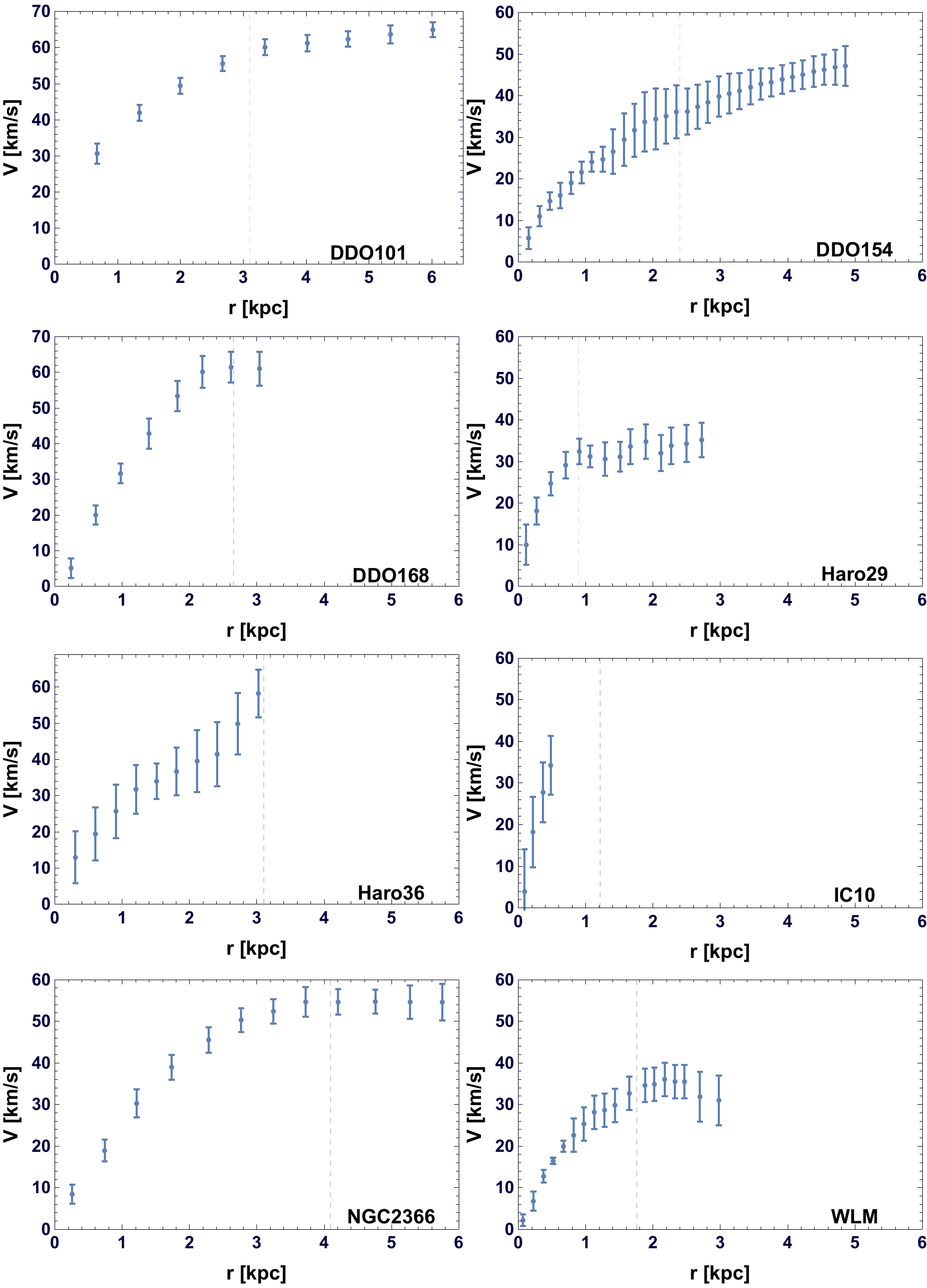}
\contcaption{}
%\label{figa1}	
\end{figure*}

 \begin{figure*}
 %\ContinuedFloat
\centering
\includegraphics[angle=0,height=18truecm,width=13truecm]{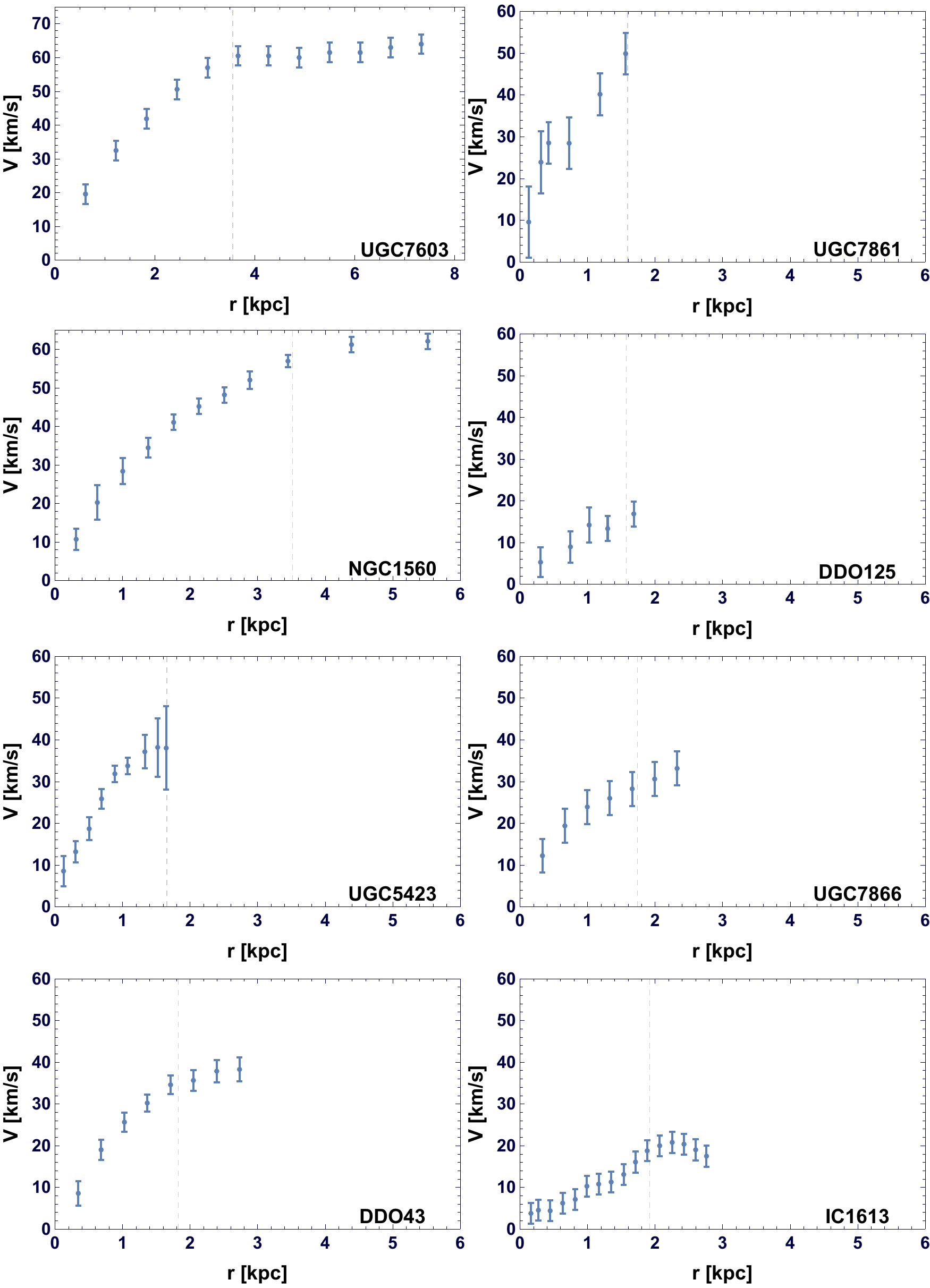}
\contcaption{}
%\label{figa1}	
\end{figure*} 

 \begin{figure*}
% \ContinuedFloat
\centering
\includegraphics[angle=0,height=18truecm,width=13truecm]{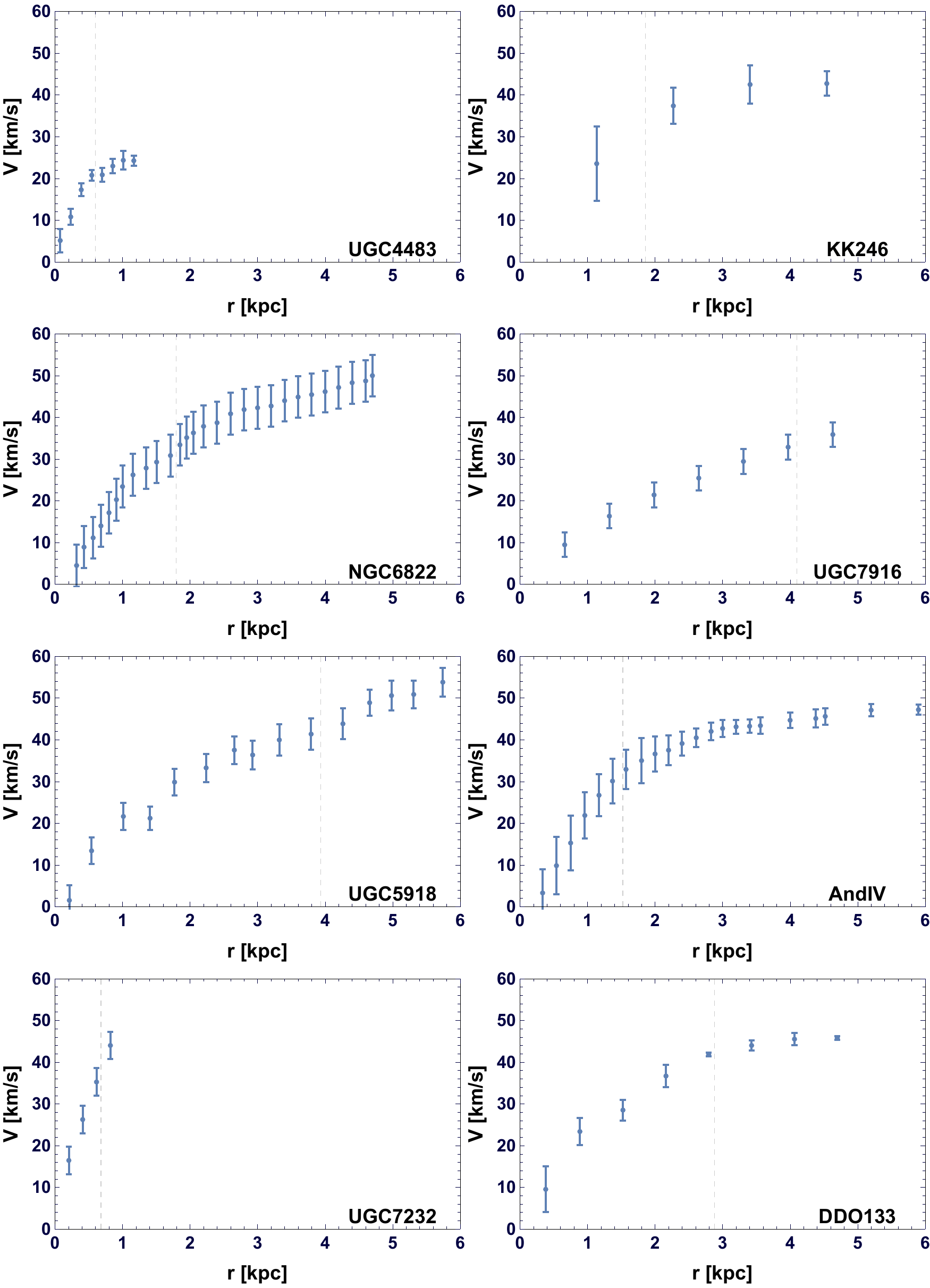}
\contcaption{}
%\label{figa1}	
\end{figure*} 
 
 \begin{figure*}
% \ContinuedFloat
\centering
\includegraphics[angle=0,height=10truecm,width=13truecm]{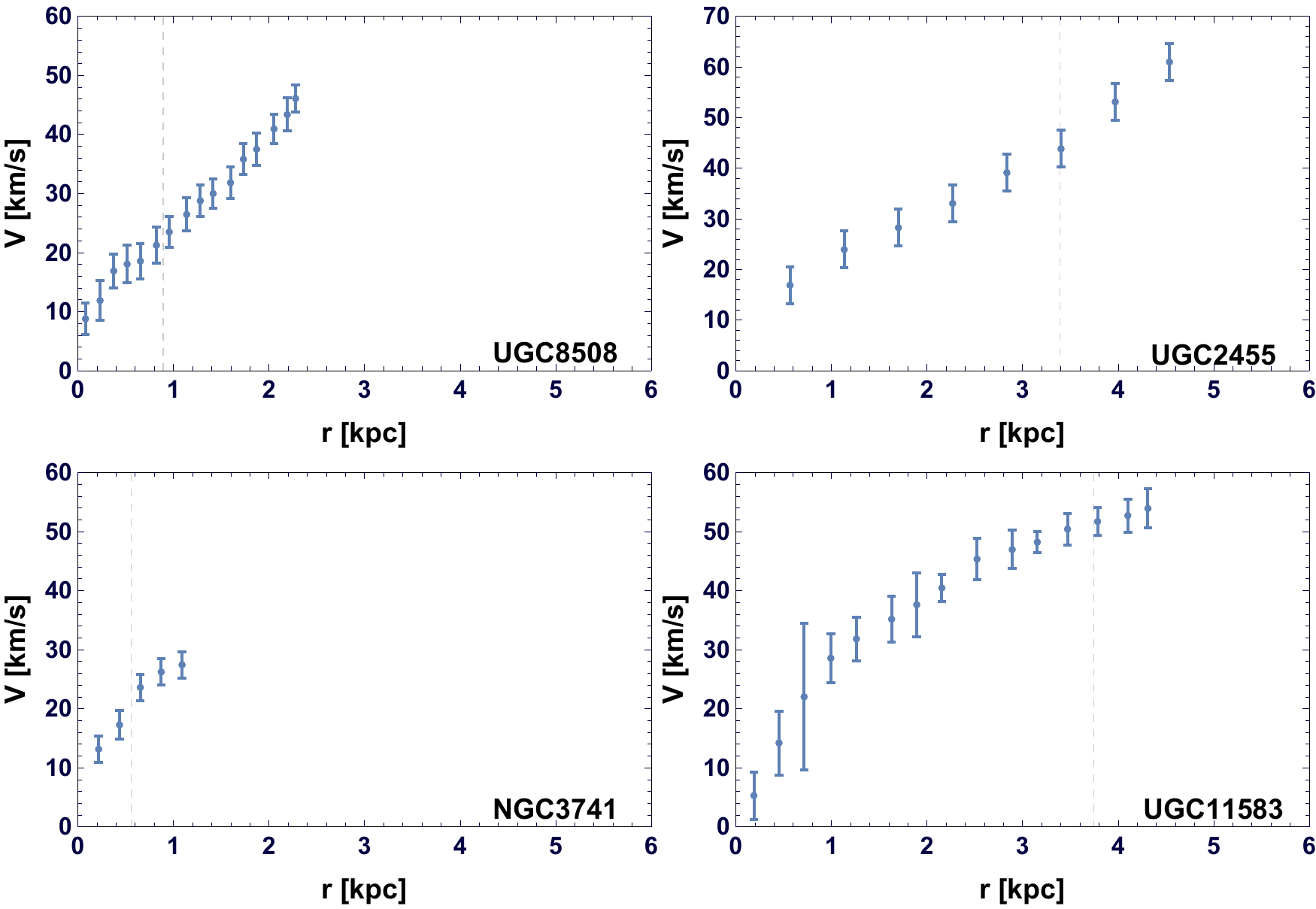}
\contcaption{}
%\label{figa1}	
\end{figure*}

\section{Comparison of the "dd"URC and the URC of PSS}

In the left panel of Fig. \ref{fig3a}, as already discussed in Section 3, we plot our coadded double normalized RC (black stars) alongside with that of the PSS's four luminosity bins. In this figure the joined red dots correspond to the 1st luminosity bin, the joined green squares correspond to the 6th luminosity bin, the joined orange diamonds to the 9th luminosity bin and the joined pink triangles correspond to the 11th luminosity bin with inside 40, 70, 40 and 16 normal spirals, respectively (see Fig. A1 of PSS). We realise that the coadded RC of the 1st PSS bin and synthetic dd RC have very similar profiles. However, the "dd"URC appears to be slightly less concentrated than that of the 1st luminosity bin of PSS. The latter might indicate the continuation
of the trend found in PSS, which states that the shape of RCs changes with luminosity. Therefore, in order to check whether we have this trend inside our sample we divided it on 3 subsamples in the following way: the most luminous, least luminous, and the middle half. Then we binned radially each subsample in the same way as described in Section 3. The data in the radial bins of each subsample are presented in Table~\ref{tbl:1app}. Furthermore, in the right panel of Fig.~ \ref{fig3a} we compare the synthetic RC of the whole dd sample with 3 synthetic RCs of the above defined subsamples. All 4 RCs agree within their uncertainties and we do not find any trend of the shape of the RCs with luminosity. However, a weak trend is impossible to reveal. In fact the small number of galaxies in each bin induce some shot noise. Therefore, in order to further investigate this we need twice as many objects.

\begin{table*}
\caption{Data in the radial bins of 3 subsamples, ordered from least to most luminous. Columns: \textbf{(1)} bin number;  \textbf{(2)} number of data points; \textbf{(3)} the central value of a bin; \textbf{(4)} 
the average coadded weighted normalized rotation velocity; \textbf{(5)} rms. on the average coadded rotation velocity.} 
%\label{tbl:2}

\centering

  \begin{tabular}{ l c c c c c c c c c c c c c } 

	\hline
	i&  N & $r_{i}$ & $v_{i}$  &$d{v_{i}}$&  N & $r_{i}$ & $v_{i}$  &$d{v_{i}}$&  N & $r_{i}$ & $v_{i}$  &$d{v_{i}}$\\

	(1) & (2) & (3) & (4) & (5) & (2) & (3) & (4) & (5) & (2) & (3) & (4) & (5) \\\hline\\
	 %\\ ---  & ---& kpc   &  km/s  & km/s  
	
 \label{tbl:1app}

     1 & 9 &  0.10  &0.23  & 0.030 & 11 & 0.11  & 0.20 & 0.025 & 11 & 0.10 & 0.20& 0.025 \\   
      2 &11 &0.25    &0.41 & 0.040 & 10 & 0.22 & 0.35 & 0.040 & 12 & 0.21& 0.38 & 0.024\\
      3 &  11 &0.41    &0.59& 0.020 & 12 & 0.35 & 0.48 & 0.033 & 12 & 0.33 & 0.56 & 0.031\\
      4 & 8  &0.55     &0.73&  0.011 & 18 & 0.51 & 0.64 & 0.027 & 17 & 0.48 & 0.70 & 0.022\\
      5 & 16  &0.70    &0.81 &  0.002 & 18 & 0.72 & 0.79 & 0.024 & 16 & 0.69 & 0.86 & 0.021 \\
      6 &  12  &0.90     &0.95 &  0.014 & 14 & 0.93 & 0.94 & 0.018 & 16 & 0.89 & 0.95 & 0.014\\
      7 &  13 &1.10    &1.01  &  0.012 & 13 & 1.11 & 1.02 & 0.013 & 10 & 1.08 & 1.01 & 0.015\\
      8 & 8 & 1.27    & 1.09 &  0.017 & 5 & 1.43 & 1.07 & 0.018 & 9 & 1.28& 1.06 & 0.032\\
      9 & 12 &1.49     &1.11  &  0.024 & 5 & 1.62 &1.08 & 0.052 & 7 & 1.49 & 1.04 & 0.015\\
      10 & 7 &1.73   &1.15 &  0.052 & 2 & 1.79 & 1.23& 0.012 & 7 & 1.80 & 1.09 & 0.025\\
      11 & 6&1.92     &1.23  &  0.056 & --- & --- & --- & --- & --- & --- & --- & ---\\
      
                     \hline
\end{tabular}
\end{table*}

 \begin{figure*}
\includegraphics[angle=0,height=9truecm,width=18truecm]{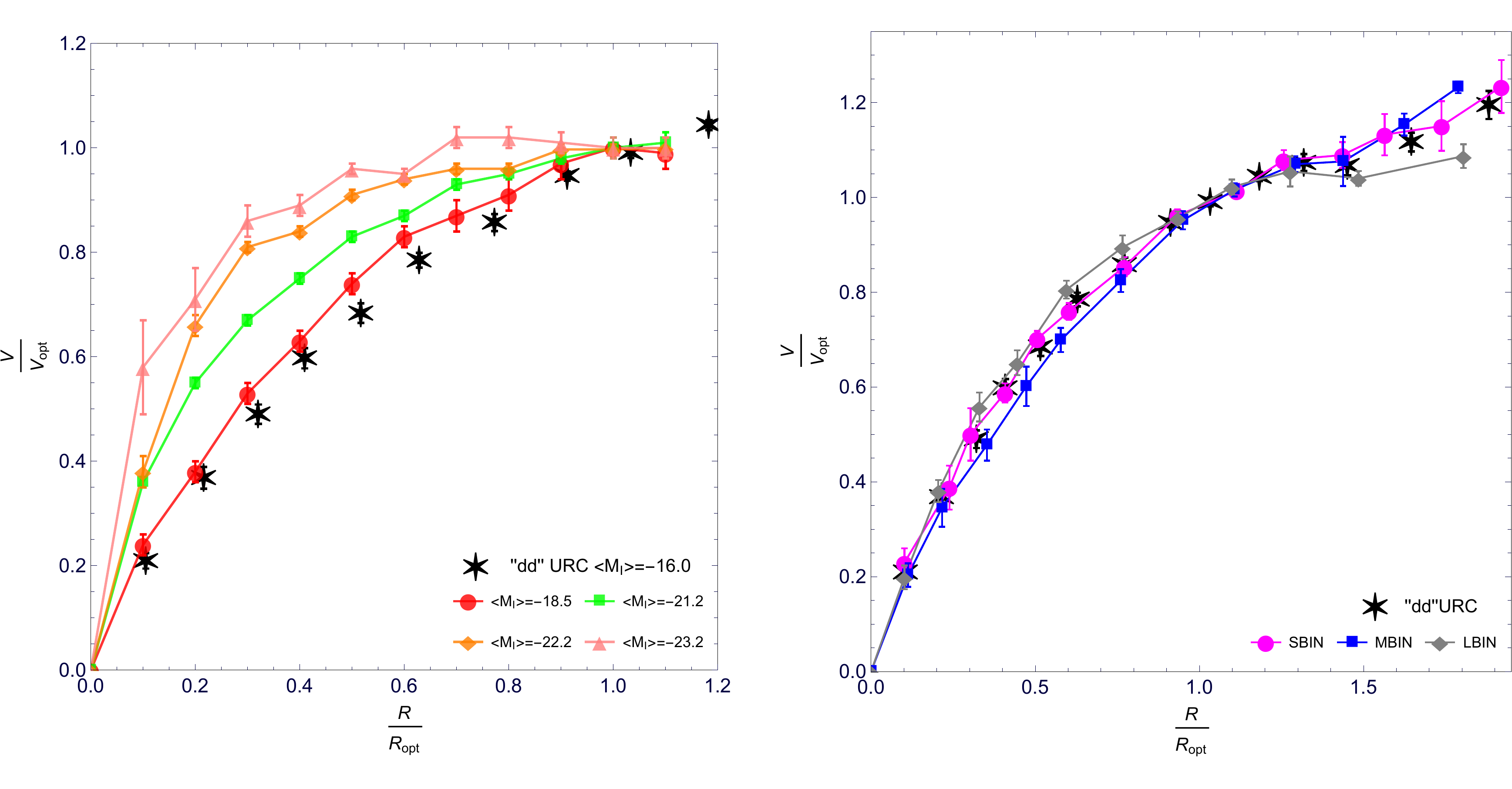}
\caption{{\it Left panel}: joined curves indicate the URC in normalized units of 4 luminosity bins of PSS. {\it Right panel}: the URC in normalized units of 3 subsamples of dwarf disc galaxies (SBIN-smallest bin; MBIN-mean bin; LBIN-largest bin). Black stars indicate the synthetic RC of the whole sample of dwarfs disc galaxies.}
\label{fig3a}	
\end{figure*}

}

\end{document}